\documentclass[conference]{IEEEtran}
\IEEEoverridecommandlockouts
\usepackage{cite}
\usepackage{amsmath,amssymb,amsfonts}
\usepackage{algorithmic}
\usepackage{graphicx}
\usepackage{textcomp}
\usepackage{lipsum}
\usepackage{physics}
\usepackage{xcolor}
\usepackage{caption}
\usepackage{subcaption}
\newtheorem{definition}{Definition}

\def\BibTeX{{\rm B\kern-.05em{\sc i\kern-.025em b}\kern-.08em
    T\kern-.1667em\lower.7ex\hbox{E}\kern-.125emX}}
\begin{document}

\title{Quantum information spreading and scrambling in a distributed quantum network: A Hasse/Lamport diagrammatic approach\\
}

\author{
\IEEEauthorblockN{Kiran Adhikari}
\IEEEauthorblockA{\textit{Institute for Communications Engineering} \\
\textit{Technical University of Munich}\\
\textit{MQV}\\
\tt kiran.adhikari@tum.de}
\and
\IEEEauthorblockN{Christian Deppe}
\IEEEauthorblockA{\textit{Institute for Communications Engineering} \\
\textit{Technical University of Munich}\\
\textit{6G-life \& MQV}\\
\tt christian.deppe@tum.de}
}

\maketitle

\begin{abstract}
Large-scale quantum networks, known as quantum internet, hold great promises for advanced distributed quantum computing and long-distance quantum communication. It is essential to have a proper theoretical analysis of the quantum network and explore new applications and protocols that justify building such an extensive network. We propose a novel diagrammatic way of visualizing information flow dynamics within the quantum network, which preserves the causal relationship between different events at different nodes. This facilitates synchronization among network nodes, studies the error propagation, and allows for tracking valuable quantum resources. Additionally, We propose a quantum information scrambling protocol, where a specific node scrambles secret quantum information across the entire network. This protocol ensures that a malicious party would need access to a significant subset of the network to retrieve the information.

\end{abstract}


\section{Introduction}

Recently, there has been a growing interest from both experimental and theoretical perspectives in building large-scale quantum networks called quantum internet \cite{wehner, qcn, doi:https://doi.org/10.1002/9781118648919.ch13, Cuomo_2020, 7562346}. Quantum internet offers exciting opportunities for advanced quantum information processing, distributed quantum computing, quantum metrology, and long-distance quantum communication \cite{Bennett_2014,singh2021quantum, PhysRevLett.67.661,K_m_r_2014, PhysRevLett.109.070503, caleffi2022distributed, yimsiriwattana2004distributed, zhang2021distributed, Buessen_2023}. Because of technological advancement \cite{pompili2021realization, daiss2021quantum, simon2017towards, humphreys2018deterministic, sangouard2011quantum}, it becomes crucial to explore these networks from the theoretical perspective as well \cite{Azuma_2021, QCNbook, van2016local,van2010distributed, avron2021quantum, qiao2022quantum, haner2021distributed, beals2013efficient, cicconetti2022resource, parekh2021quantum, PhysRevA.99.052303, chen2021review}. By studying information flows within these networks, we can effectively track the information dynamics and error propagation and ensure seamless synchronization among network nodes. At the same time, finding new protocols and applications of such distributed quantum networks justifies building a quantum internet in the first place. Therefore, the theoretical analysis of the quantum network design and the finding novel applications of such networks are of fundamental importance in large-scale quantum networks.

 Quantum information scrambling has been studied extensively in the context of quantum many-body systems \cite{r10, r11, Nahum_2018}. However, we are unaware of any connections between quantum information scrambling and distributed quantum networks. In this context, we propose a novel protocol called quantum information scrambling protocol, where the information from one node gets scrambled across the entire network such that the external malicious party cannot reconstruct it from having control of only a fraction of the nodes in the network. In Section \ref{sec:informationSpreading}, we propose a novel way of diagrammatically showing the flow of quantum information in the network. In Section \ref{sec:informationScrambling}, we present an application of a quantum network called quantum information scrambling protocol, which utilizes the diagrammatic representation of Section \ref{sec:informationSpreading}.

\subsection{Definition and conventions}

A quantum network has two fundamental elements: nodes processing quantum information and edges representing quantum channels. These nodes can be any entity able to perform local operations (LO) as dictated by the principles of quantum mechanics, for example, a client, a quantum repeater, a universal quantum computer, or an extremely powerful quantum data center (QDC) with multiple universal quantum computers.   Conversely, a quantum channel is an edge between two arbitrary nodes, allowing the exchange of a quantum system between them. Examples include an optical fiber, a superconducting microwave transmission line, or an optical free-space link.

\begin{figure}
    \centering
    \includegraphics[width= 0.5  \textwidth]{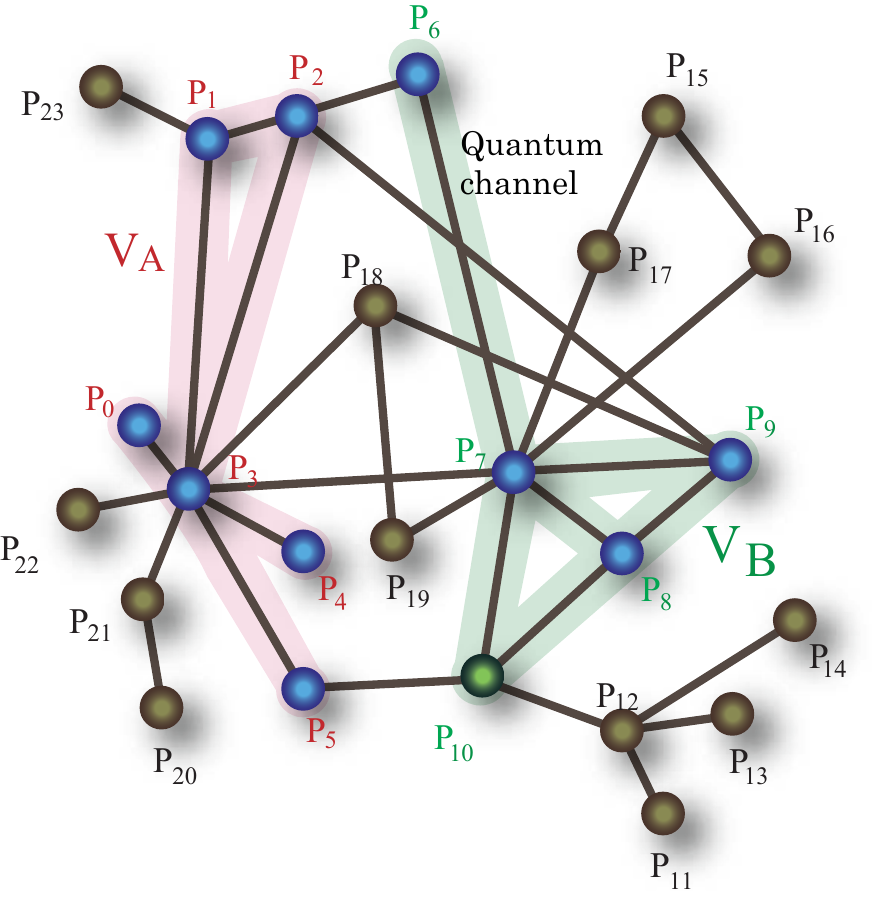}
    \caption{Quantum network with subnetworks: A network is represented by a graph $G = (V,E)$ where vertices represent quantum information processing nodes and edges represent quantum channels. A subnetwork $V_A = \{P_0, P_1, P_2, P_3, P_4, P_5 \}$ is shown with a red shade while subnetwork $V_B = \{P_6, P_7, P_8, P_9, P_{10}\}$ is represented with green shade. } 
    \label{fig:quantumNetwork}
\end{figure}

We can represent the structure of a quantum network by a graph $G = (V,E)$, as shown in Figure \ref{fig:quantumNetwork}, with a set $V$ of vertices and a set $E$ of directed edges. Vertices $V =\{P_1, P_2, .... , P_N\} $ indicates $N$ nodes with each vertex $P_i$ representing a classical or quantum information processing node. A directed edge $e = v \xrightarrow{} v'$ represents a quantum channel $\mathcal{N}^{v \xrightarrow{} v'}$.  A set of cuts $\partial V$ among the edge, at certain time $t$, can divide the quantum network into $L$ different non overlapping sub networks $V_1^t = \{P_1, ... , P_i \}, ..., V_i^t = \{P_{j+1}, ... P_k \}, ...., V_L^t = \{P_l, .... P_N\}$ where each $V_i^t$ is a subset of the graph at time $t$. Often, we will remove superscript $t$ for convenience. Furthermore, a quantum network $G$ is denoted by $G_\phi$ if every edge $e \in G_\phi$ represents a maximally entangled state. If the entanglement in a $G_\phi$ network is a free resource, we denote it by $G_\phi^\infty$.

Mathematically, a function of an arbitrary node $X$ can be represented by a local operation where, after getting quantum state $\rho$ as an input, return a quantum state $\sigma_k$ with probability $p_k$ as an output, where $\sigma_k := M_k^X \rho (M_k^X)^\dagger /p_K$  with Kraus operators $M_k^X$ satisfying $\sum_k (M_k^X)^\dagger M_k^X = 1^X$ and $p_k = \text{Tr}[(M_k^X)^\dagger M_k^X \rho ]$. Suppose there are $N$ nodes in a quantum network, and we label the $i$th node by $P_i$. A quantum state in such a network lives in a Hilbert space: $\mathcal{H} =  \mathcal{H}_1 \otimes \mathcal{H}_2 \otimes ..... \mathcal{H}_N$ where $\mathcal{H}_i \simeq \mathcal{C}^{d_i}$, $d_i < \infty$, corresponds to Hilbert space of node $P_i$. Furthermore, a quantum channel $\mathcal{N}^{X \xrightarrow{} Y}$ from a node $X$ to a node $Y$ is a completely-positive trace-preserving (CPTP) map where, $\mathcal{N}^{X \xrightarrow{} Y} (\rho) := \text{Tr}_{E'} [U^{XE \xrightarrow{} YE'} (\rho \otimes  \ket{0} \bra{0}) (U^{XE \xrightarrow{} YE'})^\dagger]$ where $U^{XE \xrightarrow{} YE'}$ is a unitary operator on  Hilbert space $\mathcal{H}^X \otimes \mathcal{H}^E$ to $\mathcal{H}^Y \otimes \mathcal{H}^{E'}$ and a state $\ket{0}^E$ of auxiliary system $E$. 

 We will also need the concept of system and subsystem sizes to use results such as the decoupling theorem \cite{Hayden_2008} of quantum information theory. The size of the node $P_i$ is denoted by  $|P_i|$, and it is related to the dimension by $d_{P_i} = 2^{|P_i|}$. If in a arbitrary protocol, each node $P_i$ dedicates $n_{P_i}$ qubits, then the size of that node would be $|P_i| = n_{P_i}$ and the dimension would be $d_{P_i} = 2^{|P_i|} =  2^{n_{P_i}}$. Then, the size of subnetwork $V_i$ is $|V_i| = \sum_{P_j \in V_i} |P_j|$, and it's dimension is $d_{V_i}  = 2^{|V_i|}$. The size of the entire network $G= (V,E)$ with $P$ different subnetworks would be  $|V| = \sum_{i=1}^P |V_i| = \sum_{i=1}^N |P_i| $  with network's dimension being $d = 2^{|V|}$. For example, if there are $n$ qubits in total participating in some arbitrary protocol in a quantum network, then the network size would be just $n$, dimension $2^n$. One can also generalize it to the case of qudits.

One primary goal of introducing the diagrammatic approach is to track various resources spent in the quantum network while running some arbitrary protocol. In a resource theoretic language \cite{r6}, $[c \to c]$ would denote the communication resource of a noiseless classical bit channel, $[q \to q]$  a noiseless qubit channel, $[cc]$ a shared, non-local bit of shared randomness and $[qq]$ a shared, noiseless EPR pair. Furthermore, other than having two-party quantum resources such as EPR pair, it is possible to have multi-party quantum resources in a network. For this, we introduce the notation $((x_1,y_1),(x_2,y_2), .. (x_i,y_i).. (x_R,y_R) )$ where $R$ indicates the number of species of resources and $x_i$ indicates the amount of $y_i$ species. Of course, the definition of species depends upon the problem in hand. For example, $((3,GHZ),(4,W))$ would imply that we have 3 GHZ states and 4 W states shared among three nodes. $R = 2$ as we have two different species of quantum resources, $GHZ$ and $W$ entangled states if we define species as three party entanglement in non equivalent way. We can also easily include information about which nodes are sharing the given resource by adding one more parameter in the notation $(x_2,y_2, P_iP_j....P_k)$. Here, $P_iP_j....P_k$ denotes the nodes where the resource $y_i$ is shared upon. 
One primary goal of introducing the diagrammatic approach is to track various resources spent in the quantum network while running some arbitrary protocol. In a resource theoretic language, $[c \to c]$ would denote the communication resource of a noiseless classical bit channel, $[q \to q]$  a noiseless qubit channel, $[cc]$ a shared, non-local bit of shared randomness and $[qq]$ a shared, noiseless EPR pair. Furthermore, other than having two-party quantum resources such as EPR pair, it is possible to have multi-party quantum resources in a network. For this, we introduce the notation $((x_1,y_1),(x_2,y_2), .. (x_i,y_i).. (x_R,y_R) )$ where $R$ indicates the number of species of resources and $x_i$ indicates the amount of $y_i$ species. Of course, the definition of species depends upon the problem at hand. For example, $((3,GHZ),(4,W))$ would imply that we have 3 GHZ states and 4 W states shared among three nodes. $R = 2$ as we have two different species of quantum resources, $GHZ$ and $W$ entangled states if we define species as three-party entanglement in non equivalent way. We can also easily include information about which nodes share the given resource by adding one more parameter in the notation $(x_2,y_2, P_iP_j....P_k)$. Here, $P_iP_j....P_k$ denotes the nodes where the resource $y_i$ is shared upon.

\subsubsection*{Quantum Information Scrambling}
The phenomenon of scrambling of quantum information has recently emerged as a fundamental concept across multiple disciplines, including many-body physics, quantum chaos, complexity theory, and black holes \cite{r10,r11, Landsman:2018jpm, Schuster:2021uvg, Sekino_2008}. Here, we present a concise overview of information scrambling while more extensive insights can be explored in the literature \cite{r10, Bhattacharyya_2022}.

\begin{figure}
    \centering
    \includegraphics[width= 0.5 \textwidth]{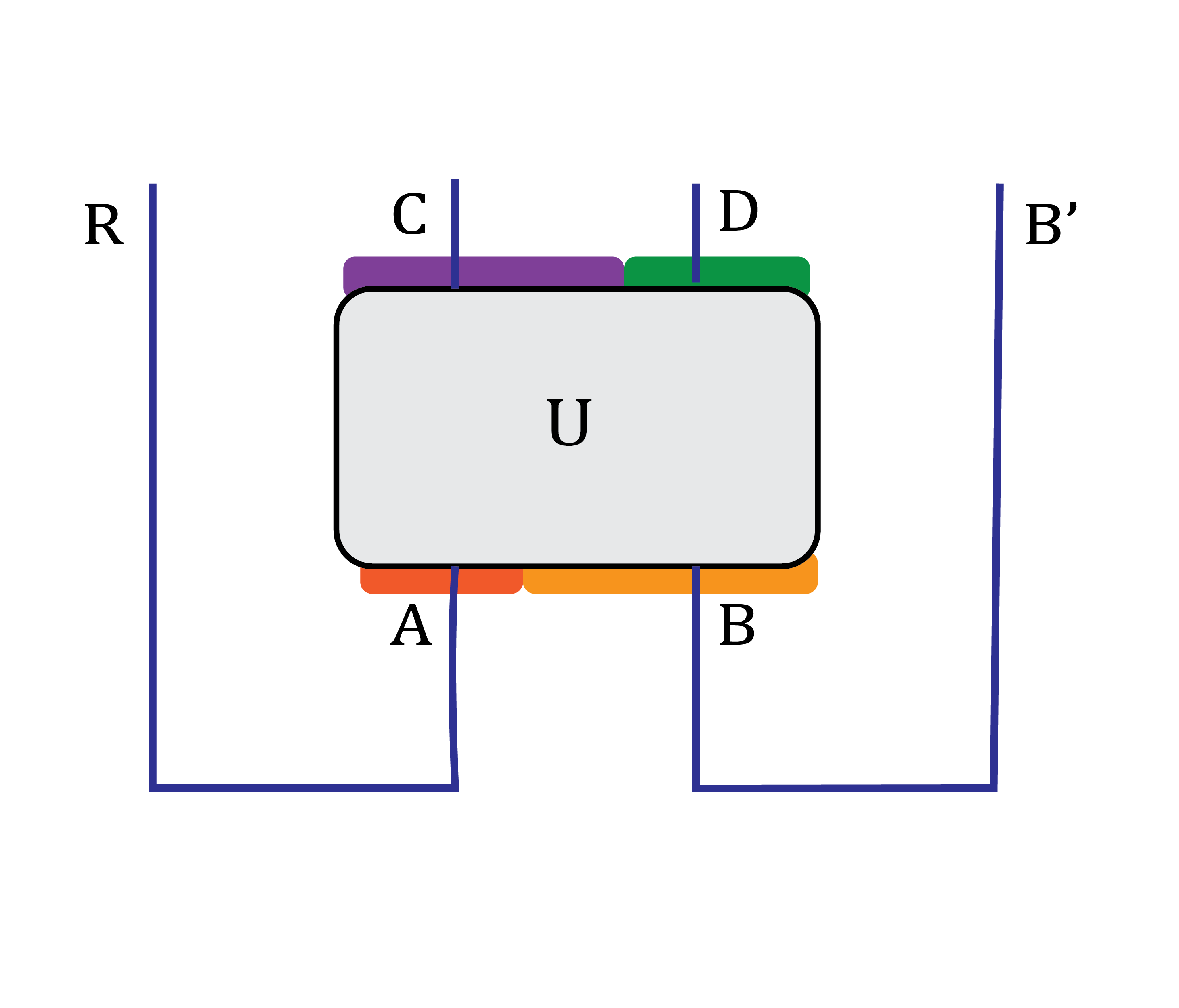}
    \caption{The quantum information of $A$ interacts with system $B$ via a scrambling unitary $U_{AB}$. System $B$ is purified by external system $B'$ while $A$ is purified by reference system $R$. Thus, the total initial state is $\ket{RA} \ket{BB'}$ while the final state after scrambling unitary $U_{AB}$ is $ \ket{\psi}_{RB'CD} = U_{AB} \otimes I_{RB'}\ket{RA} \ket{BB'}$.}
    \label{fig:informationScramblingCircuit}
\end{figure}

We can roughly understand quantum information scrambling as a phenomenon of thermalization of quantum information and is thus crucial in exploring how quantum information spreads in quantum systems. 
To grasp the essence of it, it is easier to express it using the framework of quantum circuits. Suppose we have a unitary $U_{AB}: A \otimes B \xrightarrow{} C \otimes D$, and external system $R'$ purifying $A$. The unitary is of scrambling nature if $R'$ and $C$ are almost independent for all or most  $C$ of size smaller than some parameter $l$. How one selects a subset $C$ from the composite system $CD$ becomes arbitrary as long as it remains below $l$. Mathematically, this implies that arbitrary subsystem $C$ of size less than $l$ approximately decouples from $R'$: $\rho_{R'C} \stackrel{\epsilon}{\approx} \rho_{R'} \otimes \rho_C$. Hence, the outcome of any measurement on $C$ is statistically independent of any measurement on $R'$ and can provide no information about $R'$. The parameter $l$ depends upon the purity of $B$. It is possible to change the purity of $B$ by entangling $B$ with some external system $B'$. The entire setup is depicted in Figure \ref{fig:informationScramblingCircuit}.

.

\section{Diagrammatic representation of Quantum information spreading}
\label{sec:informationSpreading}
In this section, we propose a novel way of diagrammatically representing the information flow within a quantum network motivated by the formalism of Hasse diagram  \cite{Surya_2019} and Lamport diagrams \cite{lamport}. 

Causality in a single node is trivial, as one event is the cause for the next. But, it quickly gets complicated in cooperating nodes that exchange information to solve a common task. In the realm of quantum mechanics, preserving causal relationships can be even trickier due to the presence of entanglement. Nonetheless, despite this intricacy, a causal framework should persist due to the no-signaling theorem \cite{r6}, which states the impossibility of sending information just using entanglement alone. Furthermore, if we have sufficient non-local quantum resources like entanglement, all non-local quantum computations can be shifted to local quantum computations and classical communication. With this, most multi-party quantum protocols can be phrased as a synchronization problem among different nodes.

Considering the large scale of quantum networks, which may comprise numerous nodes with many qubits each, creating a quantum circuit diagram for the entire network is impractical.
Instead, this diagrammatic approach focuses on maintaining a casual relationship across events in different nodes, thus providing a robust tracking mechanism for information dynamics and error propagation within the network. This approach also facilitates the seamless synchronization of different nodes within the network while tracking resource expenses, thereby enabling the successful implementation of quantum protocols.

\subsection{Space-time diagram}
The concept of the Hasse/Lamport diagram has found prior applications within the realm of classical computing (see \cite{garg2002elements, attiya2004distributed, kshemkalyani2011distributed, thulasiraman2011graphs}). Furthermore, it has been employed in physics, particularly in specific quantum gravity models (see \cite{rideout1999classical, Surya_2019}). In this work, we use the Hasse/Lamport diagrammatic approach to visually depict quantum protocols. 

\begin{figure*}
    \centering
         \includegraphics[width= 0.9 \textwidth]{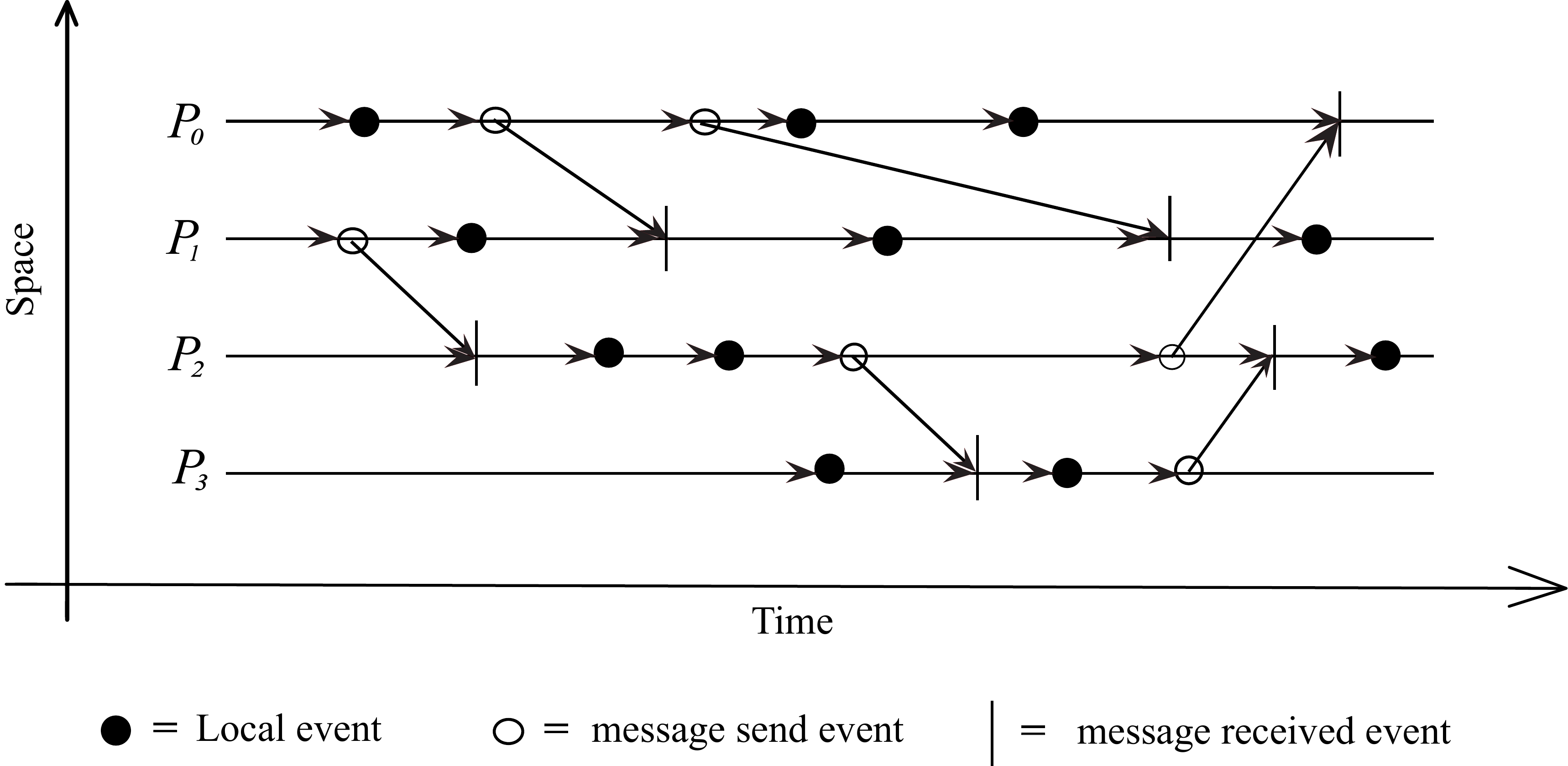}
         \caption{Example of a space-time diagram with four nodes. The dots represents local event while arrows represents message send event.}
         \label{fig:spaceTimeDiagram}
\end{figure*}

The evolution of distributed execution is given by a Hasse space-time diagram where each node $P_i$ has a corresponding horizontal line that tracks the progress of that node. 
 The binary relationship arrow indicates the casual relationship from one event to another. A solid dot can mark a local event. A line at the end of the arrow can indicate the message received event. Messages can be sent via quantum or classical channels. Because of the Hasse-like structure, the diagram has interesting properties. Suppose $x$, $y$, and $z$ indicate three events, and if there is an arrow connecting between two events, we indicate it by a binary relation $\prec$ with properties: 
\begin{enumerate}
    \item Acyclic: x $\prec$ y and y $\prec$ x $\Rightarrow$ x = y
    \item Transitive: y $\prec$ y and y $\prec$ z $\Rightarrow$ x $\prec$ z 
\end{enumerate}
The transitive property guarantees that we don't have a loop in the diagram, which otherwise would imply a future affecting the past and violate causality. Space-time diagram \ref{fig:spaceTimeDiagram} shows an example of a space-time diagram with four nodes. In the following sections, we will only use the arrow in the message send event for convenience and drop the line at the message received event.

\subsection{Local event}
Defining an event in a quantum network can be trickier because of non-local resources such as entanglement and the requirement to track both quantum and classical states. In this section, we propose a novel definition of a local event based on the notion of density matrix and its change. The quantum state of a node $P_i$ can be mathematically described by a local density matrix $\rho_i$ obtained by tracing out the rest of the network, $\rho_i = \Tr_{\{V/P_i \}}(\rho)$, where $\rho$ is the quantum state of the entire network. This characterizes both the classical and quantum state of the node $P_i$. We say a local event occurred at node $i$ whenever this local (reduced) density matrix $\rho_i$ changes. The change in the density matrix can be computed using 
$L_1$ norm  which for any operator $M$ is defined as $||M||_1 = \text{Tr} \sqrt{M^\dagger M}$. With this, we propose a following definition of a local event.

\begin{definition}[Local event]
A local event is recorded at node $P_i$ at time $t'$ if $||\rho_i(t') - \rho_i(t'')||_1 \geq \epsilon$, where $t''$ is the time of the immediate previous event at the same node $P_i$. The threshold parameter $\epsilon$ depends upon the protocol and its requirements. 
\end{definition}

This definition is also physically motivated because  for $||\rho_i(t') - \rho_i(t'') ||_1 < \epsilon$, $\text{Tr}(\Pi (\rho_i(t') -  \rho_i(t''))) < \epsilon$ for any projection operator $\Pi$. This implies that the probability outcome of any experiment between two density matrices $\rho_i(t')$ and $ \rho_i(t'')$ differs by at most $\epsilon$ when $||\rho_i(t') - \rho_i(t'') ||_1 < \epsilon$. Therefore, if the density matrix doesn't change up to a certain threshold, operationally, it is unnecessary to call it an event as no new information can be obtained from it.

\begin{figure}
    \centering
    \includegraphics{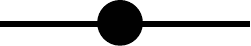}
    \caption{Local event denoted by a dot}
    \label{fig:localNode}
\end{figure}

This way of defining a local event is also ideal from the perspective of causality. Suppose two nodes $P_i$ and $P_j$ share a bipartite system and are even allowed to be entangled. No signaling theorem or, in general, no communication theorem \cite{r6} of quantum mechanics says that nothing $P_i$ chooses to do using local computations on his side will have any effect on the local density matrix of $P_j$. With this definition of a local event, we can then guarantee that the event at one node cannot affect other nodes unless the information is transferred via a quantum or classical channel. This serves as a foundation for maintaining causality in the space-time diagram. A future of an event $x$ denotes all events casually affected by $x$, and the past of an event $x$ are all the events that could have affected $x$. In the space-time diagram \ref{fig:localNode}, a dot represents a local event at that node.

\subsection{Quantum and Classical Communication}
Based on the definition of a local event, for one node to casually affect the other node, it must use a communication channel. The communication in a quantum network between two nodes can be via quantum or classical channel \cite{howe2023robust}. 

\begin{figure*}
    \centering
         \includegraphics[width = \textwidth]{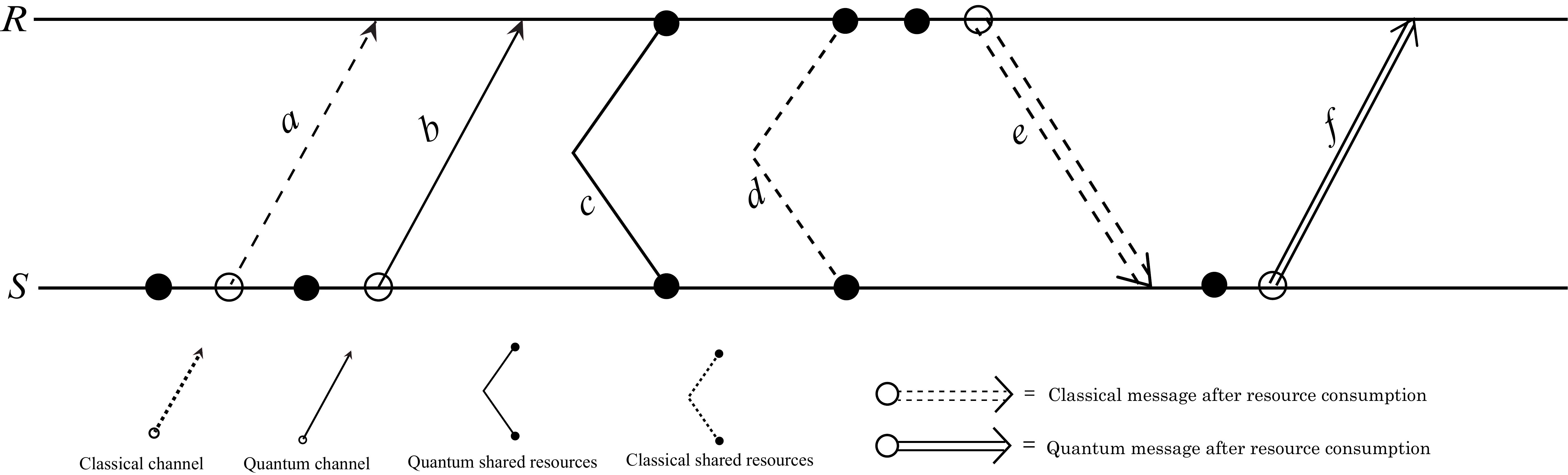}
         \caption{Space-time diagram with the following sequences: local event at node S represented by a solid dot, classical communication from S to R via dotted arrow, local event at S, quantum communication from S to R, shared quantum resources between R and S, shared quantum resources between R and S, local event at R, classical communication from R to S following resource consumption, local event at S, and then quantum message from S to R after the resource consumption. The  labels $a$, $b$, $c$, $d$, $e$, and $f$ indicates the amount of unit resources. For example, $a$ and $e$ could indicates number of bits, while $b$ and $f$ qubits. Similarly, $c$ indicates amount of maximally entangled qubits between $R$ and $S$ and $d$ indicates the amount of shared classical resources.}
         \label{fig:quantumCommunication}
\end{figure*}

In the space-time diagram, quantum processes are typically represented with a solid line, while classical processes are typically represented with a dotted line. Furthermore, usually, non-local computations are not allowed. In the space-time diagram depicted in Figure \ref{fig:quantumCommunication}, we represent a message transfer through a classical channel using a dotted slant arrow while a message transfer through a quantum channel using a solid slant arrow. The labels indicate the unit message, which could, for example, be bit or qubit. This allows one to track the resource consumption and generation in the network.

\subsection{Shared resources}

A fundamental way a quantum network differs from a classical one is by introducing new resources, such as quantum entanglement, which cannot be achieved via classical resources alone. We represent classical and quantum resources using dotted and solid curves rather than arrows. The shared classical resource between two nodes, R and S, is denoted by a dotted curved shape ending with dots at each node as shown in Figure \ref{fig:quantumCommunication}. In contrast, a solid curve denotes shared quantum resources. The label in the diagram denotes the amount of shared resources. For example, it could be the amount of classical common randomness or entanglement. Because a maximally entangled pair is a fundamentally important resource in a quantum network, we represent it by a solid angle shape rather than a curve, terminating with dots at each node. Similarly, we can also represent shared common randomness by a similar shape but a dotted one.


\subsection{Resource generation}

Using space-time diagrams, we will now show how to generate resources between two nodes. Because it is a shared resource, local processes cannot generate it alone. Therefore, it requires a communication channel, albeit quantum or classical. Furthermore, quantum resources cannot be generated via classical channel communication alone. 

\begin{figure*}
     \centering
     \begin{subfigure}[b]{0.4\textwidth}
         \centering
         \includegraphics[width=\textwidth]{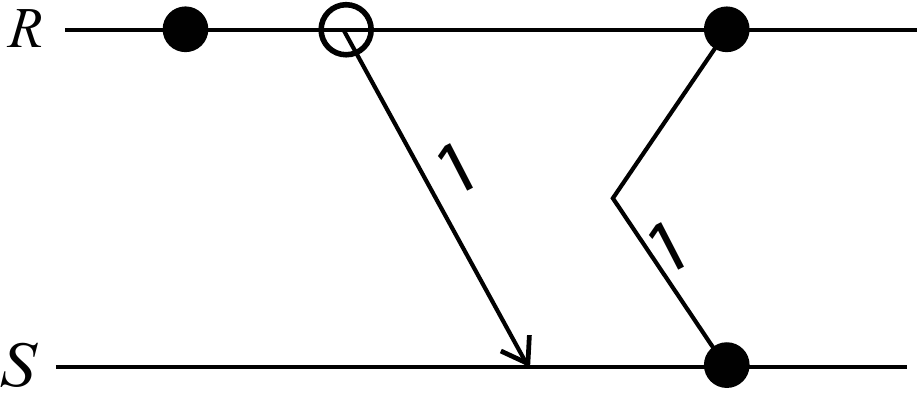}
         \caption{ }
         \label{fig:re1}
     \end{subfigure}
     \hfill
     \begin{subfigure}[b]{0.4\textwidth}
         \centering
         \includegraphics[width=\textwidth]{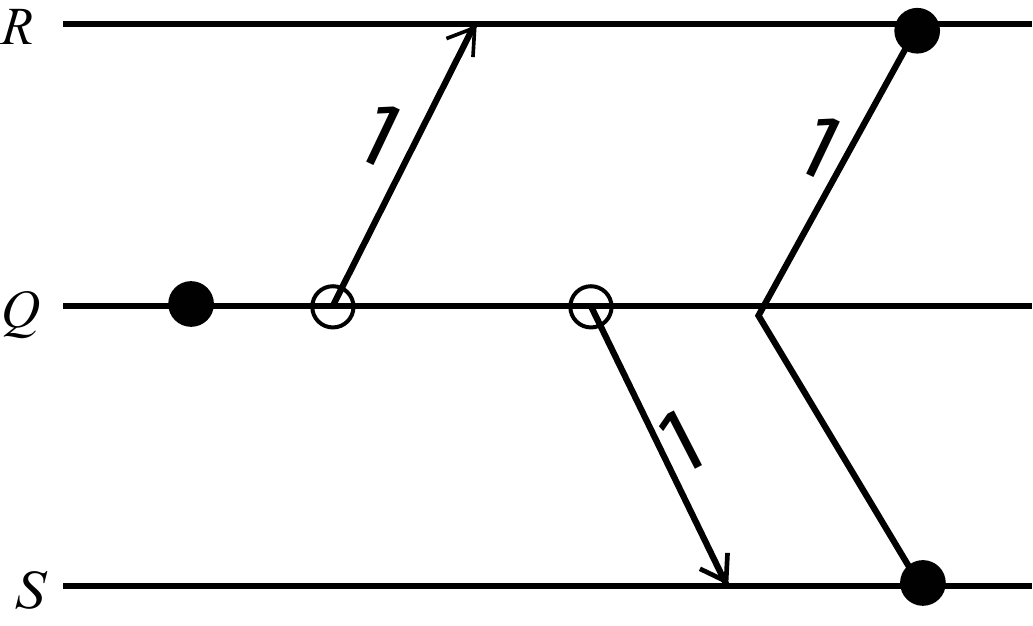}
         \caption{ }
         \label{fig:re2}
     \end{subfigure}
     \hfill
          \begin{subfigure}[b]{0.4\textwidth}
         \centering
         \includegraphics[width=\textwidth]{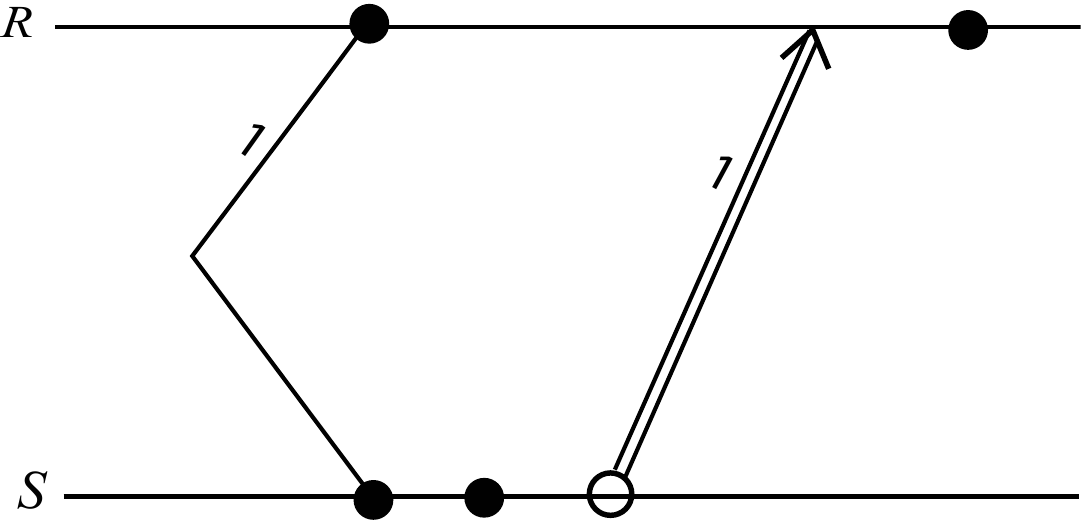}
         \caption{ }
         \label{fig:re3}
     \end{subfigure}
     \hfill
     \begin{subfigure}[b]{0.5\textwidth}
         \centering
         \includegraphics[width=\textwidth]{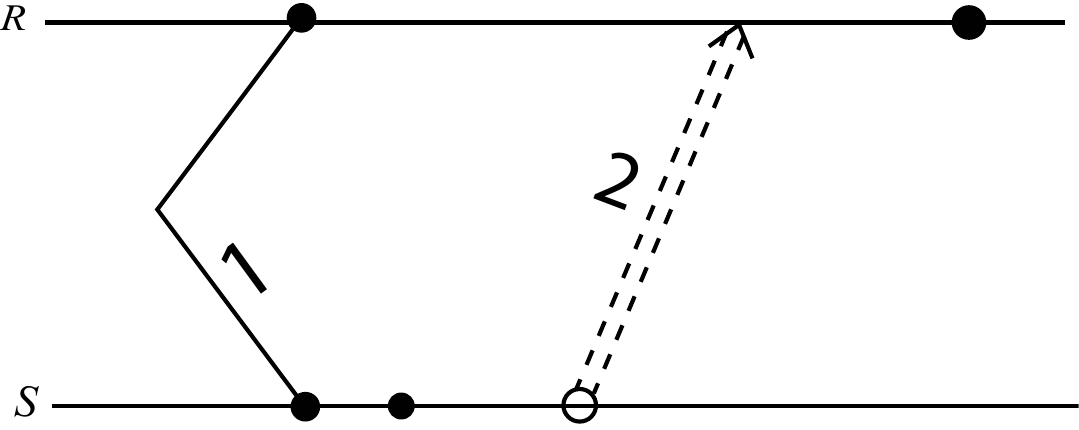}
         \caption{ }
         \label{fig:re4}
     \end{subfigure}
        \caption{Resource generation and consumption. (\textbf{a}) and (\textbf{b}) shows resource generation among R and S.  (\textbf{c}) and (\textbf{d}) shows two fundamental protocol, superdense coding and quantum teleportation respectively, that can be implemented by consuming these resources. }
        \label{fig:resources}
\end{figure*}

As an example, let us consider two ways entanglement can be generated between nodes R and S.  The first approach entails R preparing a maximally entangled pair and transmitting one of the qubits from the pair to node S through a quantum channel, as shown in the space-time diagram \ref{fig:re1}. In the second approach, a third node Q generates a maximally entangled pair and sends one qubit from the pair to node R and another to S, which results in the establishment of a shared quantum resource between nodes R and S. This approach is depicted in space-time diagram \ref{fig:re2}.


\subsection{Resource consumption}

The exciting thing about quantum resources is that, by consuming them, one can do nontrivial tasks that would have otherwise been impossible. Therefore, it is necessary to distinguish between usual communication and communication after resource consumption. In a space-time diagram,  we can represent a message sent after consuming quantum or classical resources using a classical channel by a dotted double-line arrow. We can represent the usage of the quantum channel after consuming these resources by a solid double-line arrow. as shown in the space-time diagram \ref{fig:quantumCommunication}. The double line here doesn't imply that we are sending two units of messages. Instead, the label in the space-time diagram will tell how many units of messages are transferred. For example, we will show two fundamental protocols in a space-time diagram, which consumes quantum entanglement to achieve specific nontrivial tasks.


\subsubsection{Superdense coding}
Super dense coding protocol uses pre-shared entangled qubits to encode two classical bits of information by transmission of a single qubit:
\[[qq] + [q \rightarrow q] \geq 2 [c \rightarrow c] \]
In the space-time diagram \ref{fig:re3}, R and S share a pair of EPR pairs. S performs an encoding on her half of the EPR pair and then sends it to R via a quantum channel denoted by a double-line solid arrow. Note that even though S sends a single qubit, it is represented by a double-line arrow in a space-time diagram.

\subsubsection{Teleportation}

Quantum teleportation protocol uses pre-shared entanglement to transfer a quantum state from one node to another via classical communication only
\[[qq]+2[c \to c] \geq [q \to q]. \] The protocol is shown in the space-time diagram \ref{fig:re4} where dotted double line arrow is to indicate usage of classical channel after consuming quantum resources. In the space-time diagram \ref{fig:re4}, the label $2$ indicates sending two classical bits of messages.


\subsection{Multiparty case}

\begin{figure*}
     \centering
     \begin{subfigure}[b]{0.4\textwidth}
         \centering
         \includegraphics[width=\textwidth]{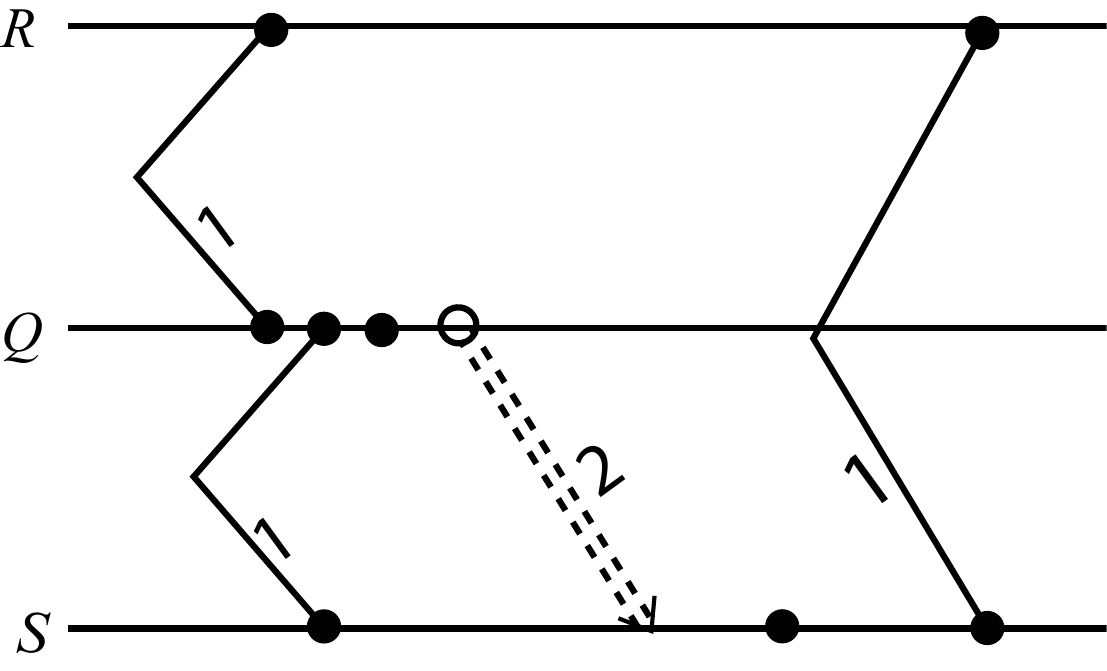}
         \caption{ }
         \label{fig:entanglementSwapping1}
     \end{subfigure}
     \hfill
     \begin{subfigure}[b]{0.5\textwidth}
         \centering
         \includegraphics[width=\textwidth]{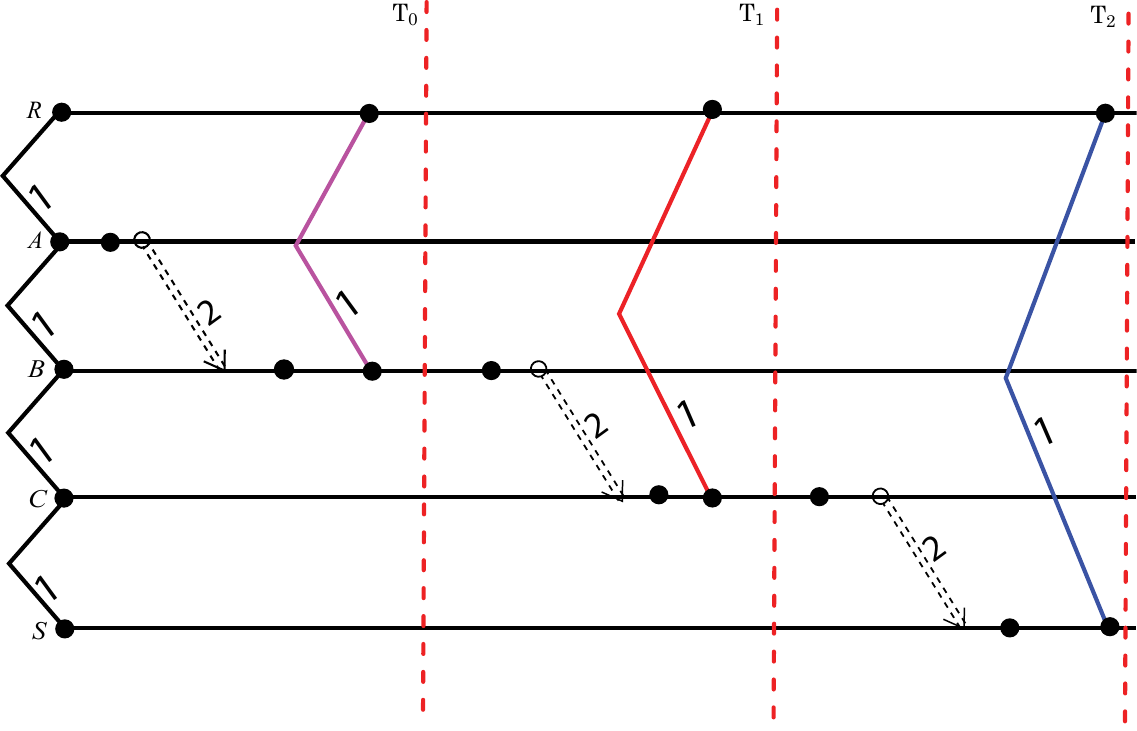}
         \caption{ }
         \label{fig:entanglementSwapping2}
     \end{subfigure}
      \hfill
     \begin{subfigure}[b]{0.5\textwidth}
         \centering
         \includegraphics[width=\textwidth]{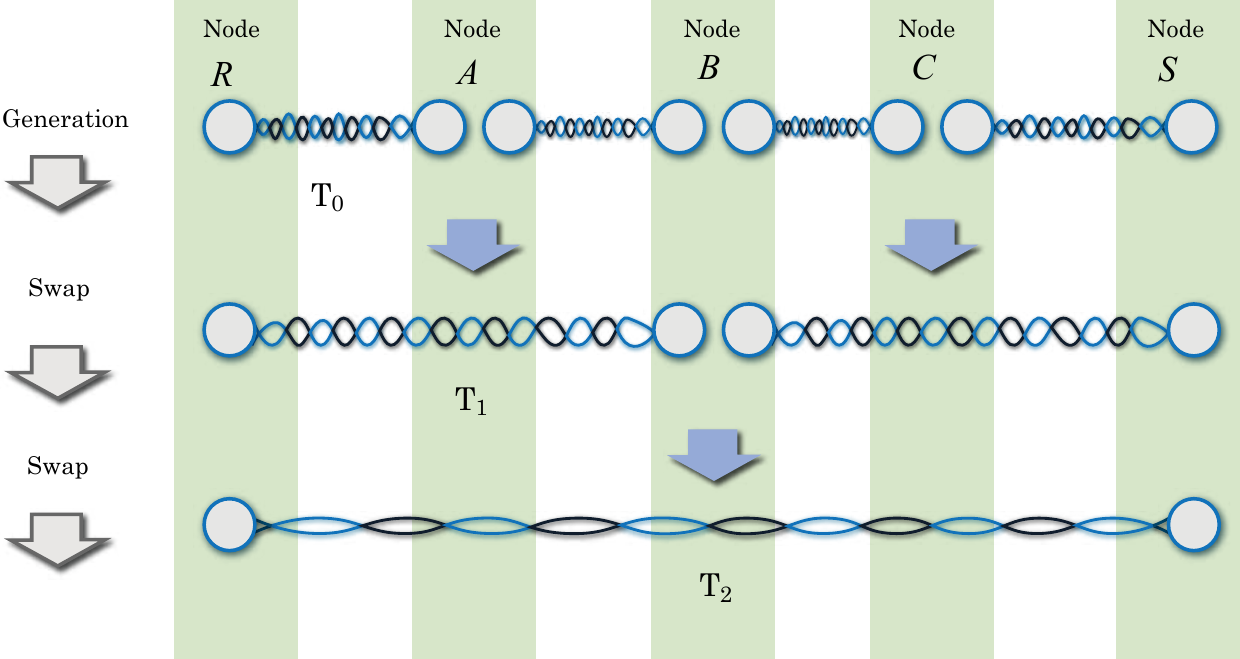}
         \caption{ }
         \label{fig:entanglementSwapping3}
     \end{subfigure}
        \caption{ Entanglement swapping protocol used for building quantum repeater (\textbf{a}) Node R and Node Q are maximally entangled, while Node Q and Node S are maximally entangled. The goal is to get Node R and Node S maximally entangled.  Node Q performs a local computation and sends classical information to Node  S. This results in Node R and Node S being maximally entangled.  (\textbf{b}) Space-time diagram showing entanglement swapping protocol among 5 nodes in three time steps. (\textbf{c}) Entanglement swapping protocol illustrated in a conventional literature style }
        \label{fig:entanglementSwapping}
\end{figure*}

The true potential of these space-time diagrams arises for the multiparty case. For two-party cases or a network with relatively few nodes, quantum circuits could suffice and might even offer an advantage. However, for the multiparty scenario, with a large number of nodes and a large number of processes, the quantum circuit description would be highly impractical. Nonetheless, space-time diagrams can offer a means to extract numerous essential features of such complicated network protocols. 

In the space-time diagram \ref{fig:entanglementSwapping1}, we've illustrated a crucial three-party protocol used in quantum repeater, called entanglement swapping, involving three nodes Q, R, and S. At the end of the protocol, one can observe that R and S are entangled by checking the dots on the end of a curved line. Moving to the space-time diagram depicted in Figure \ref{fig:entanglementSwapping2}, we've expanded the scenario to involve more nodes and presented in Figure \ref{fig:entanglementSwapping3}, a commonly used illustration in the literature to depict the same concept.

One can have multiparty classical and quantum resources for the multiparty case, which we denote by a repeated curved $\mathcal{W}$ shape with a hinge touching each participating node. Again, the shape will be dotted for classical resources and solid for quantum. We can use the following notation to label the resource $((x_1,y_1),(x_2,y_2), .. (x_i,y_i).. (x_N,y_N) )$, where $N$ indicates the number of species of resources and $x_i$ indicates the amount of $y_i$ species. We can also use this notation $(x_2,y_2, P_iP_j....P_k)$ to label the resources to indicate which nodes are participating. However, this information can also be obtained from the space-time diagram itself. 

\begin{figure}
    \centering
         \includegraphics[width=0.4\textwidth]{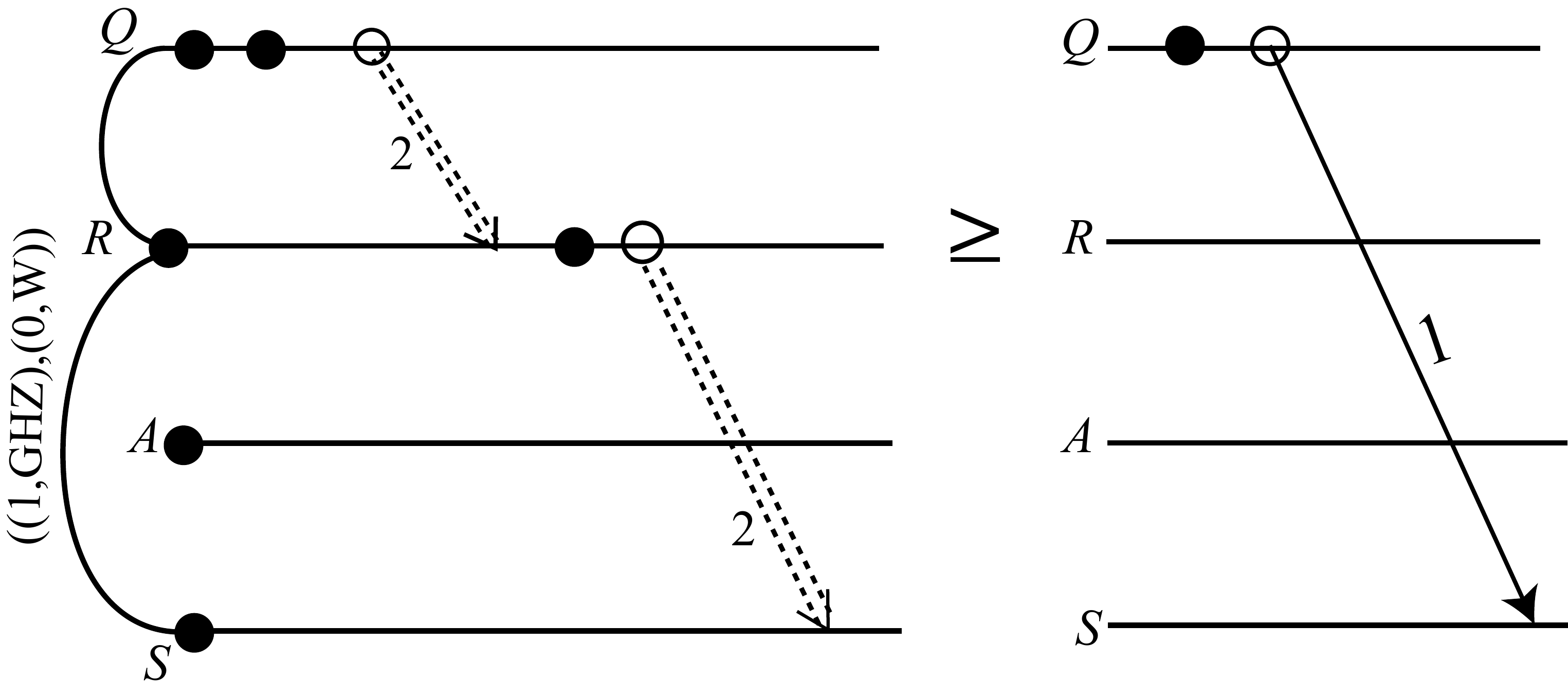}
         \caption{Space-time diagram showing third party controlled teleportation from node Q to node S using GHZ states and only classical communication. The resource inequality is: 
$[qqq]_{QRS} + 2 [c \to c]_{Q \to R} + 2 [c \to c]_{R \to S} \geq [q \to q]_{Q \to S} $
where $[qqq]_{QRS}$ indicates the GHZ state shared between Q, R and S. }
         \label{fig:thirdPartyTeleportation}
\end{figure}

For example, in the space-time diagram for a three-party case, it is possible to have two different families of shared quantum resources, such as GHZ or W entangled states. Suppose we label it as $((3, GHZ),(4, W))$, then we have 3 GHZ states and 4 W states shared among three nodes. Here, we refer to GHZ and W states as two different families as they are not equivalent under Local Operation and Classical Communication (LOCC); however, how one categorizes different types of resources depends on their definition. The three-party controlled teleportation, a variation of teleportation protocol, is shown in the space-time diagram \ref{fig:thirdPartyTeleportation} to demonstrate this multiparty resources and consumption concept.

\begin{figure*}
     \centering
     \begin{subfigure}[b]{0.4\textwidth}
         \centering
         \includegraphics[width=\textwidth]{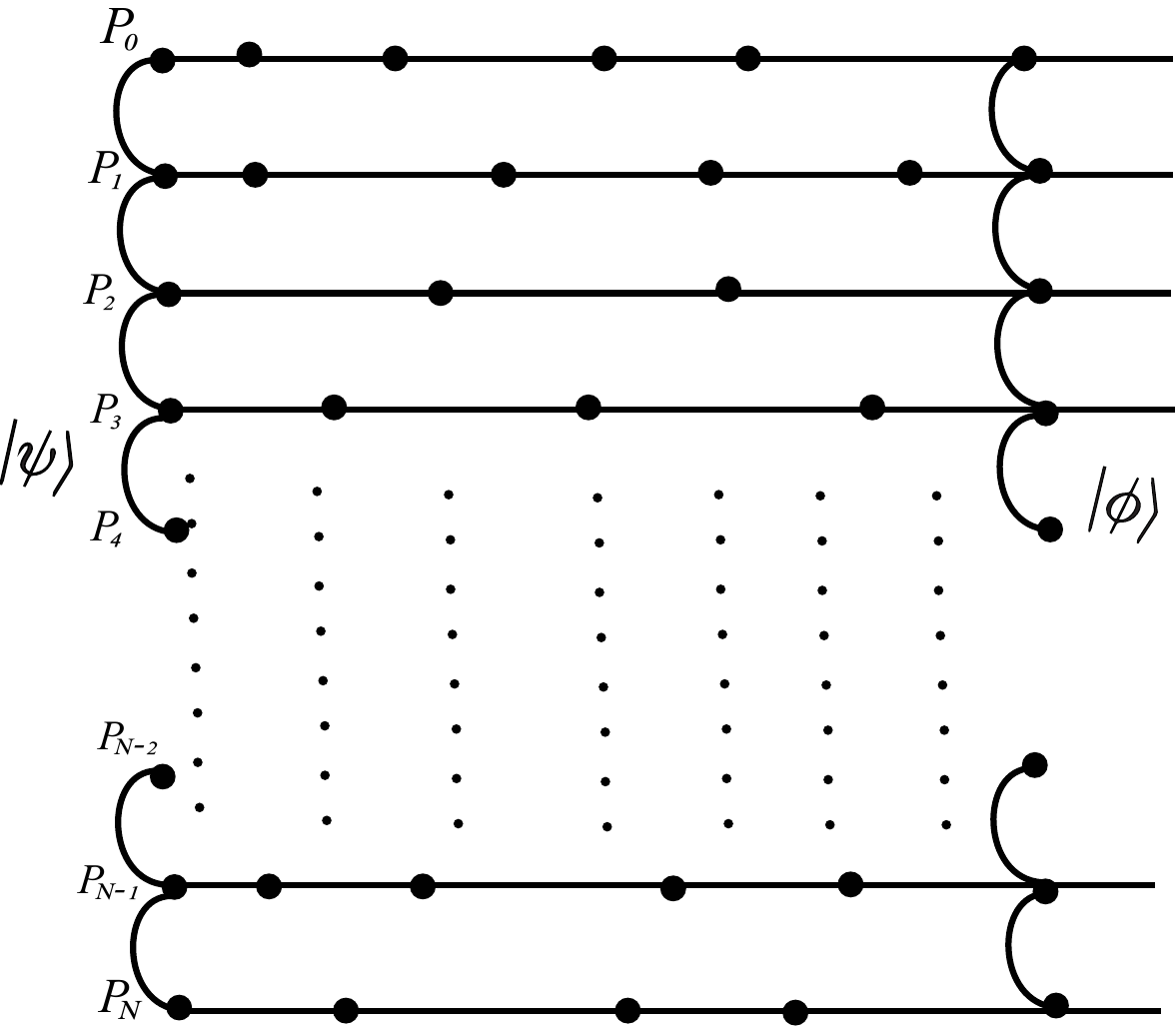}
         \caption{ }
         \label{fig:LU}
     \end{subfigure}
     \hfill
     \begin{subfigure}[b]{0.4\textwidth}
         \centering
         \includegraphics[width=\textwidth]{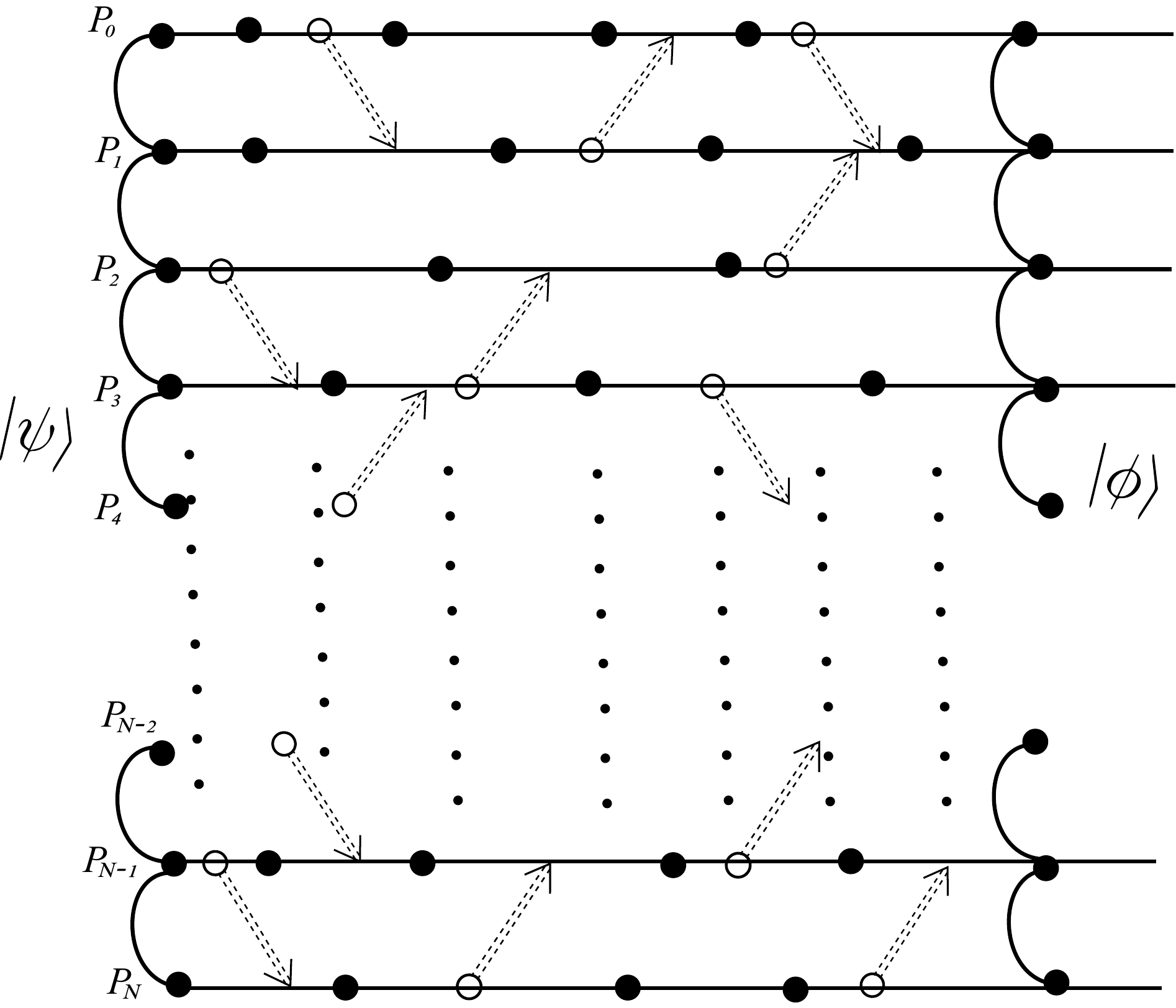}
         \caption{ }
         \label{fig:LOCC}
     \end{subfigure}

        \caption{ (\textbf{a}) LU protocol where local nodes perform local operation but no communication among nodes is allowed. (\textbf{b}) LOCC protocol where nodes can perform local operation but only classical communication is allowed among nodes.  In the subsequent round, another node may perform an operation depending on measurement from other nodes. The label $\ket{\psi}$ and $\ket{\phi}$ denote initial and final multiparty quantum state indicated by curved line touching multiple nodes.}
        \label{fig:LULOCC}
\end{figure*}

There are two important protocols called local unitary equivalence (LU) and equivalence under local operations and classical communication (LOCC) equivalence, which doesn't require the use of a quantum channel.  Two state vectors are considered equivalently entangled under LU if they differ only by a local unitary basis.
\begin{equation}
    \ket{\psi} \equiv_{\text{LU}} \ket{\phi} \xrightarrow{} \ket{\psi} = (U_1 \otimes .... \otimes U_N)  \ket{\phi}
\end{equation}
Figure \ref{fig:LU} shows the space-time diagram for such a protocol.
Similarly, we say that two states are considered LOCC-equivalent if they can be transformed into each other through a protocol that only involves the usage of a classical channel as shown in space-time diagram \ref{fig:LOCC}.

\subsection{Noise and types of operations}
\begin{figure}
    \centering
         \includegraphics[width=0.3\textwidth]{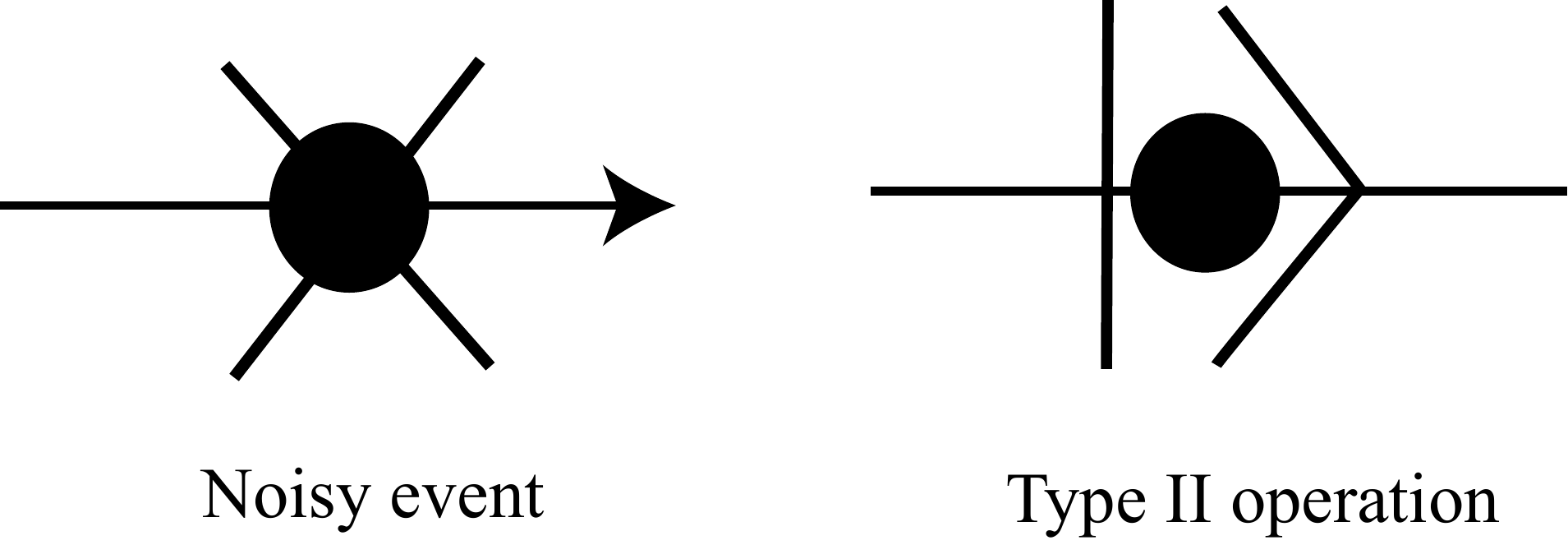}
         \caption{Noisy event or communication can be indicated by a cross  while type II operation can be indicated by a dot inside a Ket notation. }
         \label{fig:noise}
\end{figure}

One crucial advantage of a space-time diagram is that it helps to analyze the error propagation in a quantum network. We can denote a noisy local operation and communication with a cross, as shown in Figure \ref{fig:noise}. 

So far, we have assumed that all local operations are equally powerful. However, it is unrealistic, especially for distributed quantum networks, where some nodes might be more powerful. One can introduce different types of local operations based on their protocol. One way of classifying local operations would be based on the complexity the node can handle. For example, one can have type I and type II operations, where type I can only implement stabilizer protocol \cite{Veitch_2014} while type II could be a fully universal quantum computation. In the space-time diagram, as shown in figure \ref{fig:noise}, a dot inside a Ket notation represents a local process with a Type II operation, and a dot represents a Type I operation.

\subsection{Examples}

\begin{figure*}
\label{fig:applications}
     \centering
     \begin{subfigure}[b]{0.3\textwidth}
         \centering
         \includegraphics[width=\textwidth]{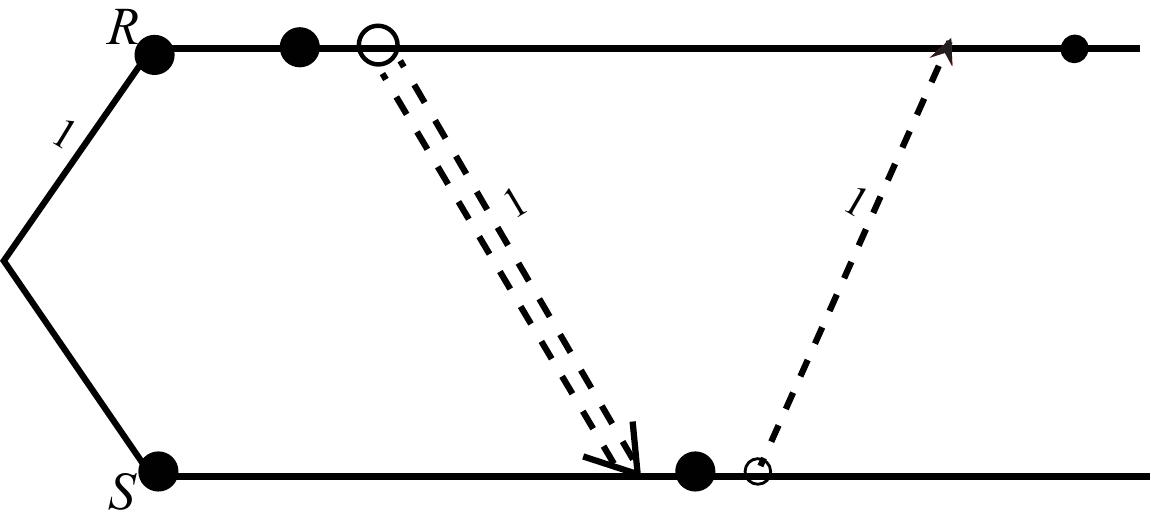}
         \caption{Distributed CNOT }
         \label{fig:distCNOT}
     \end{subfigure}
     \hfill
     \begin{subfigure}[b]{0.8\textwidth}
         \centering
         \includegraphics[width=\textwidth]{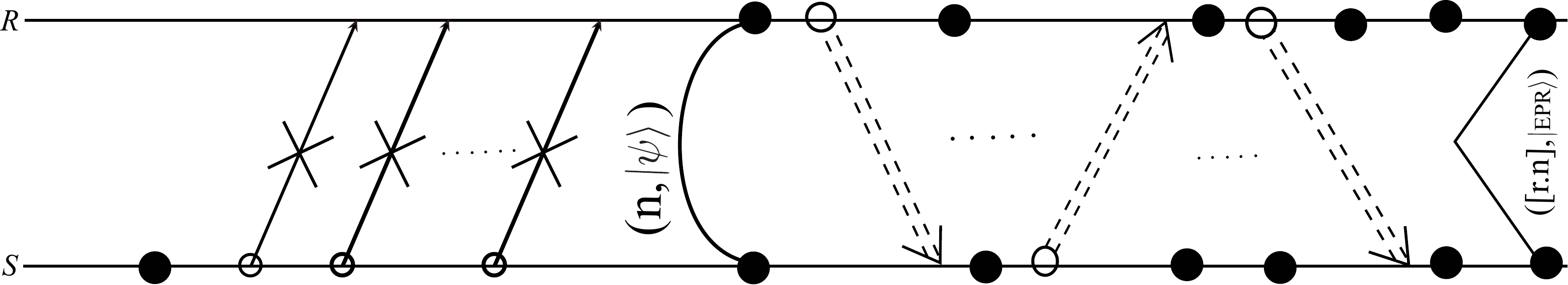}
         \caption{Entanglement distillation}
         \label{fig:EntanglementDistillation}
     \end{subfigure}
      \hfill
     \begin{subfigure}[b]{0.4\textwidth}
         \centering
         \includegraphics[width=\textwidth]{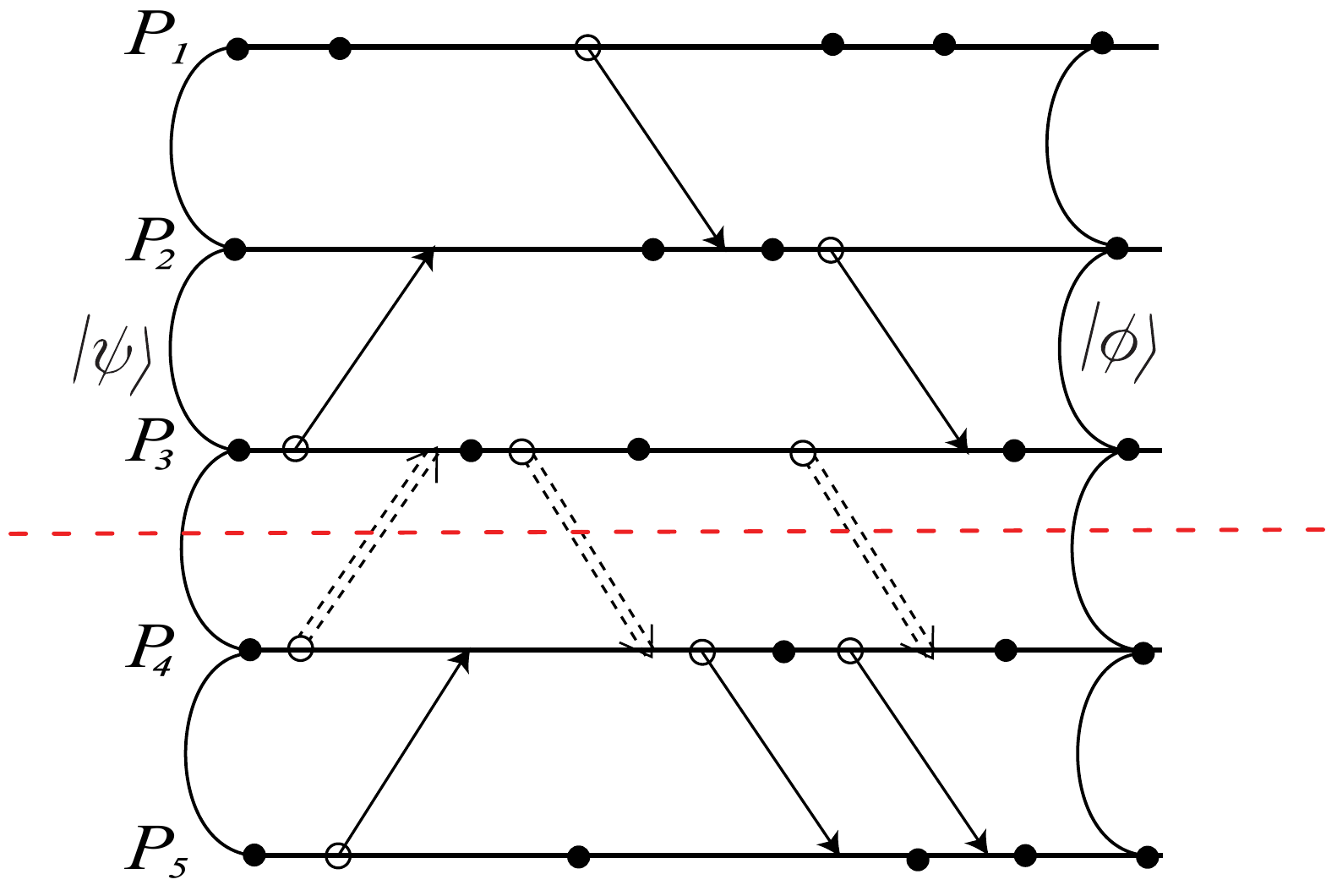}
         \caption{Nielsen theorem}
         \label{fig:NielsenThereom}
     \end{subfigure}

        \caption{(\textbf{a}) Space-time diagram for the distributed CNOT gate as proposed in \cite{PhysRevA.62.052317}.
        (\textbf{b}) Entanglement distillation protocol using only LOCC $    \ket{\psi}^{\otimes n} \xrightarrow{} \ket{\text{EPR}}^{\otimes [r.n]}$.
(\textbf{c}) Space-time diagram showing the diagrammatic representation of Nielsen's theorem \cite{PhysRevLett.83.436}. Nielsen's theorem gives a simple criterion for the equivalence of bipartite pure states under LOCC. In the diagram, the network of 5 nodes $V = \{P_1, P_2, P_3, P_4,P_5 \}$ is divided into two subnetworks $V_1 = \{P_1, P_2, P_3 \}, V_2 = \{ P_4,P_5\}$ making it a bipartite state. Within subnetworks $V_1$ and $V_2$, they are allowed to do quantum communications, but across the $V_1$ and $V_2$, they are only allowed to do classical communication. Nielsen's theorem gives a criteria on how final state $\ket{\phi}$ can be reached starting from initial state $\ket{\psi}_{V_1V_2}$.}
        \label{fig:three graphs}
\end{figure*}

In this section, we will show some examples of how space-time diagrams can be used to represent quantum protocols. The first example is of a distributed CNOT gate as shown in the space-time diagram \ref{fig:distCNOT}. The second example is of entanglement distillation (see figure \ref{fig:EntanglementDistillation}), a LOCC protocol, where out of $n$ copies of bipartite pure system $\psi$, one can generate copies of EPR pair at a rate $r$. The third example (see figure \ref{fig:NielsenThereom}) is a protocol that demonstrates Nielsen's theorem, a simple criterion for the equivalence of bipartite pure states under LOCC.

\section{Quantum information scrambling protocol in a quantum network}

\label{sec:informationScrambling}

\begin{figure*}
     \centering
     \begin{subfigure}[b]{0.3\textwidth}
         \centering
         \includegraphics[width=\textwidth]{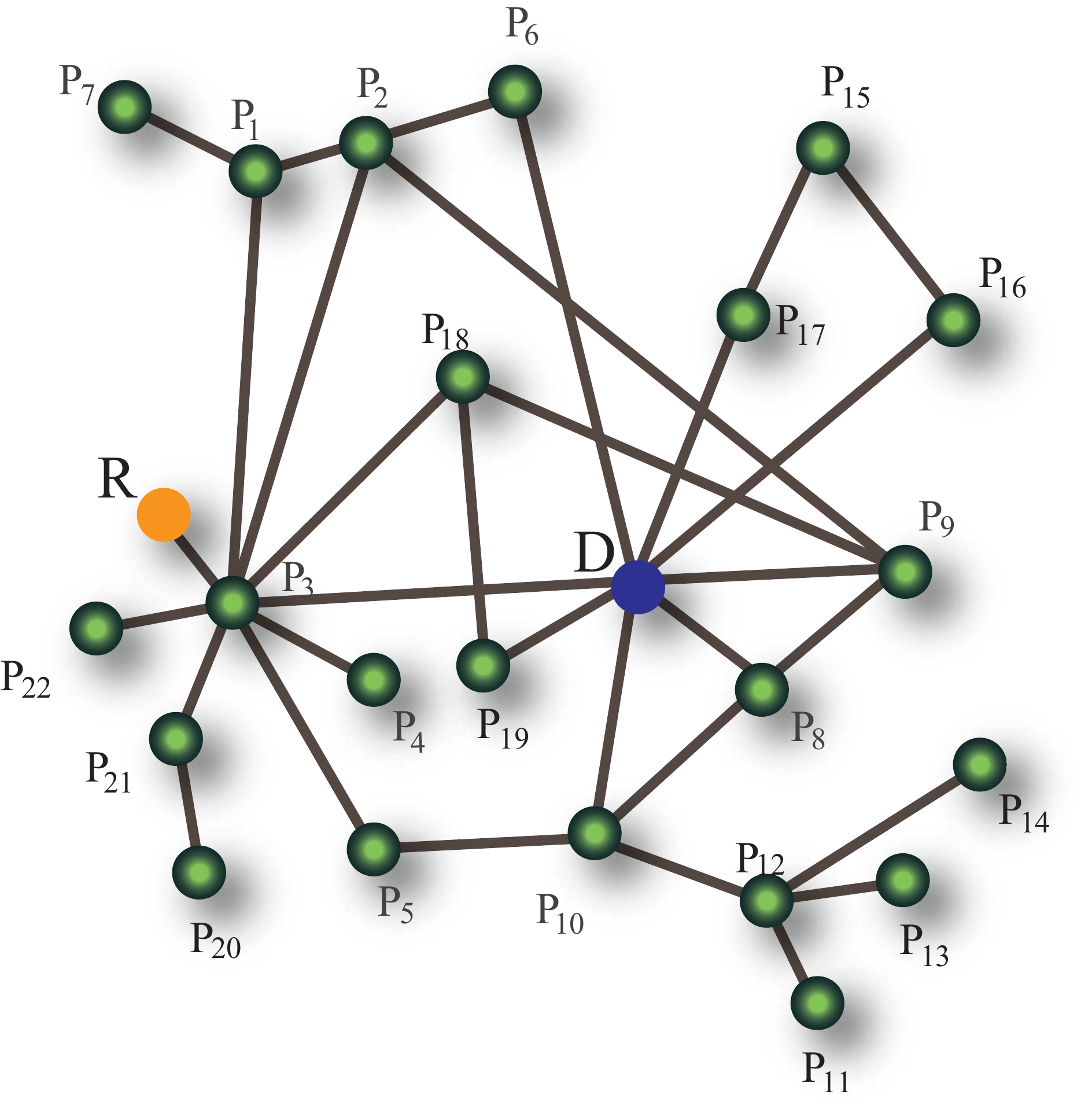}
         \caption{A node $R$, marked yellow, wants to scramble its quantum information across the quantum network. $D$ denotes the quantum data center node.}
     \end{subfigure}
     \hfill
      \begin{subfigure}[b]{0.3\textwidth}
         \centering
         \includegraphics[width=\textwidth]{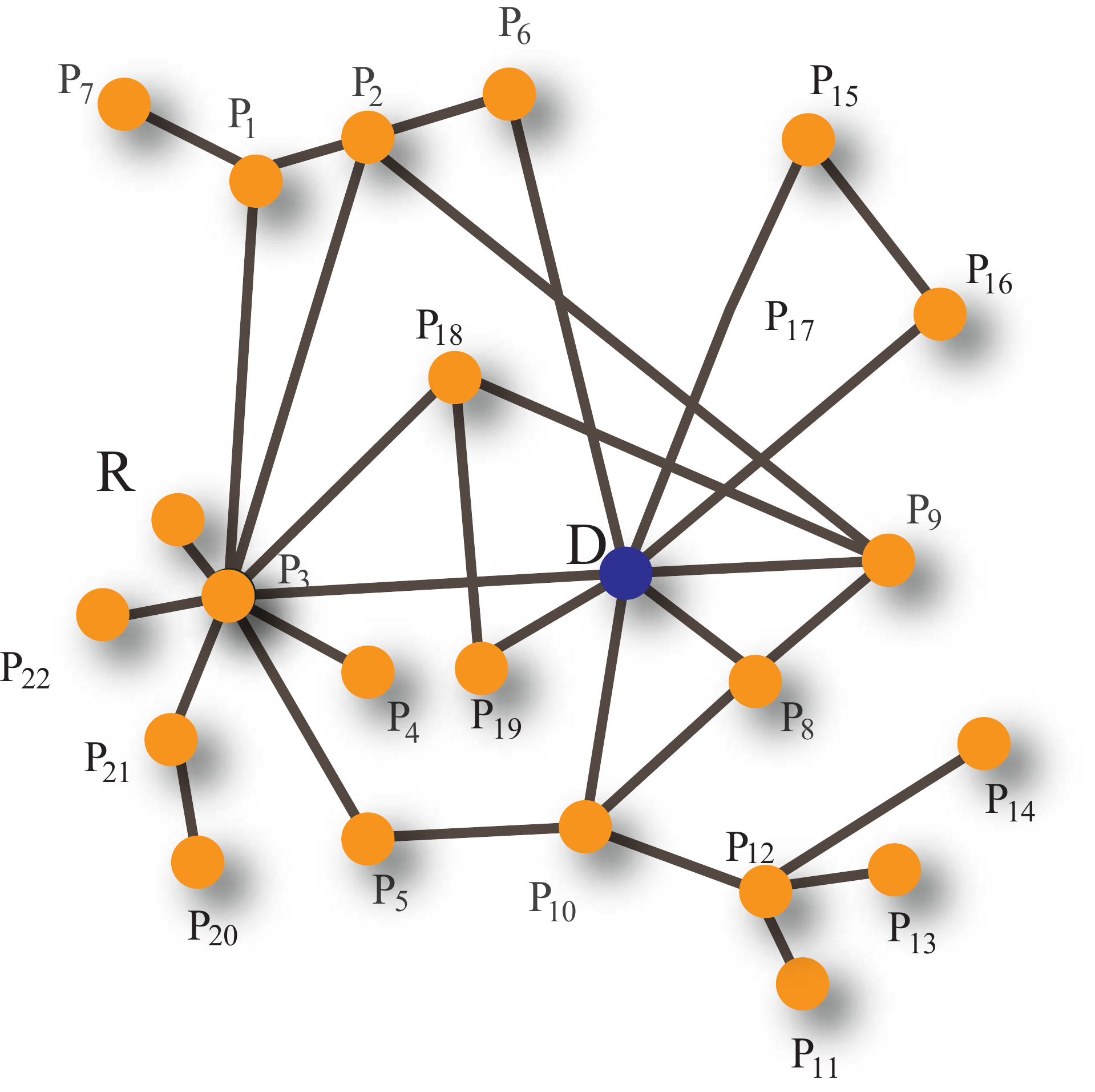}
         \caption{Network after the information scrambling protocol. }
     \end{subfigure}
     \hfill
     \begin{subfigure}[b]{0.3\textwidth}
         \centering
         \includegraphics[width=\textwidth]{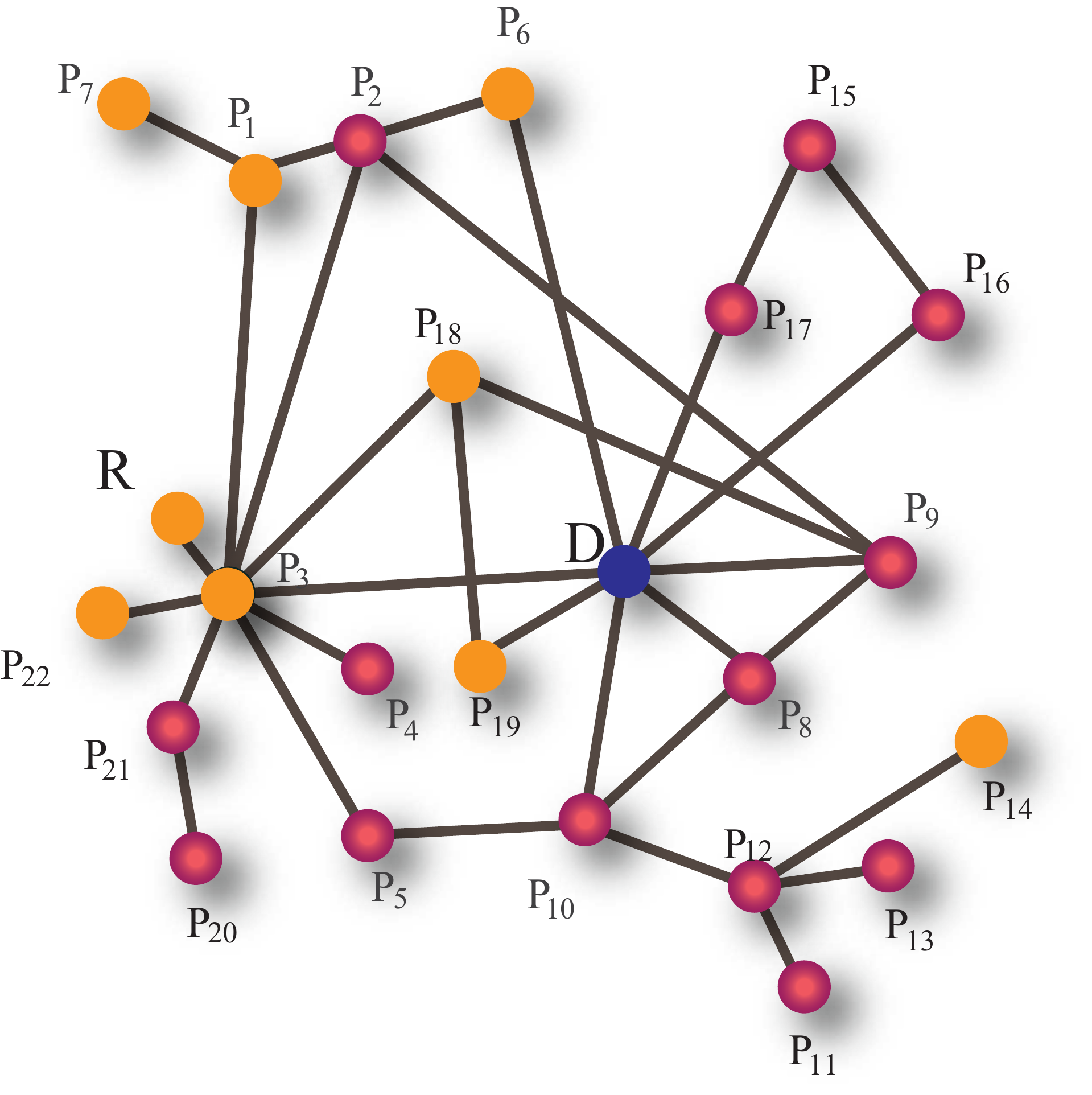}
         \caption{Network after the malicious party $E$, marked by red, gets access to a subset of the network. }
     \end{subfigure}
     \hfill
     \begin{subfigure}[b]{0.7\textwidth}
         \centering
         \includegraphics[width=\textwidth]{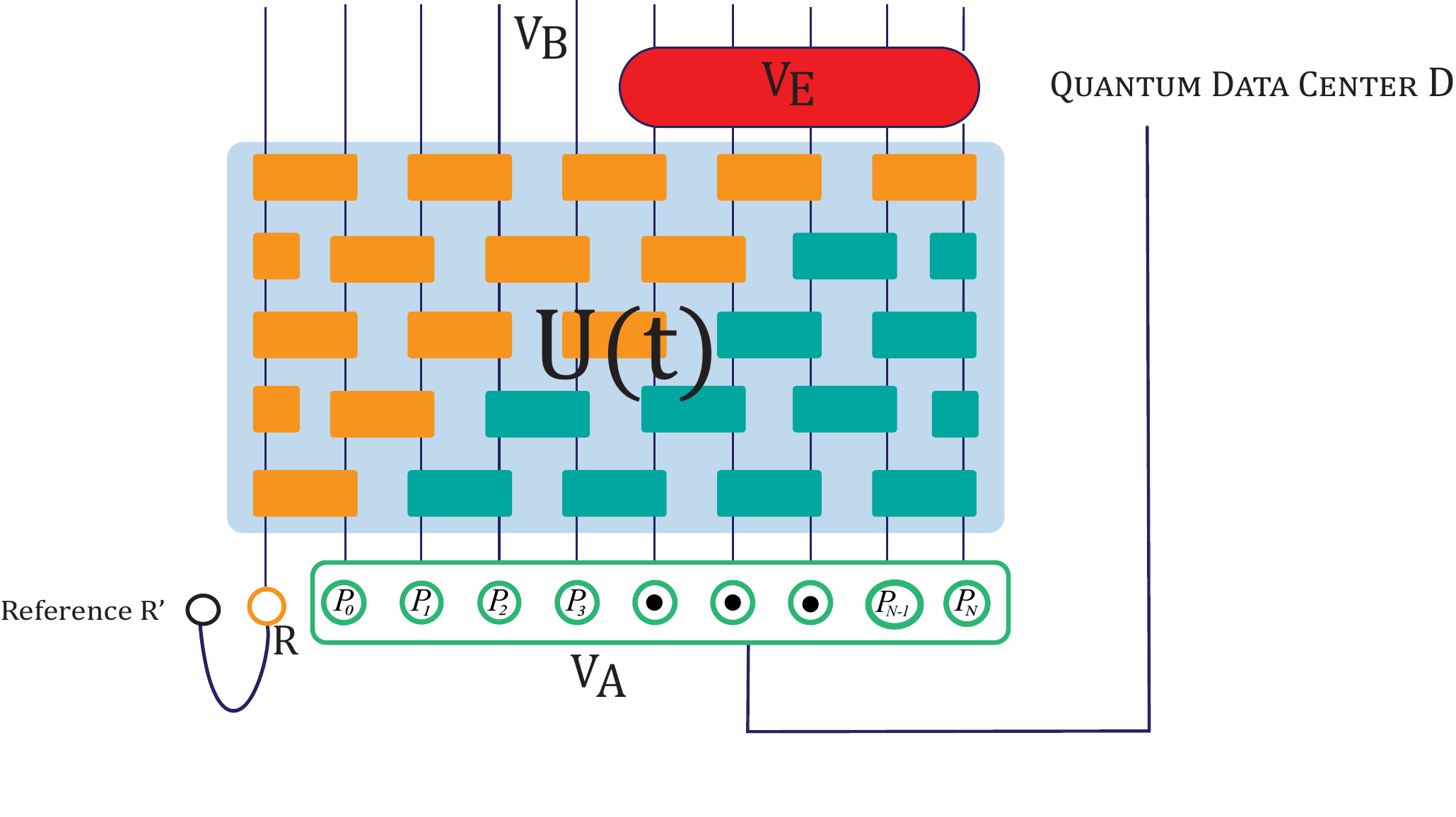}
         \caption{ A circuit representation of information scrambling protocol similar to Figure \ref{fig:informationScramblingCircuit}. The external system R' is entangled with R, which helps track the network's information dynamics. }
     \end{subfigure}
     \hfill
     \begin{subfigure}[b]{0.6\textwidth}
         \centering
         \includegraphics[width=\textwidth]{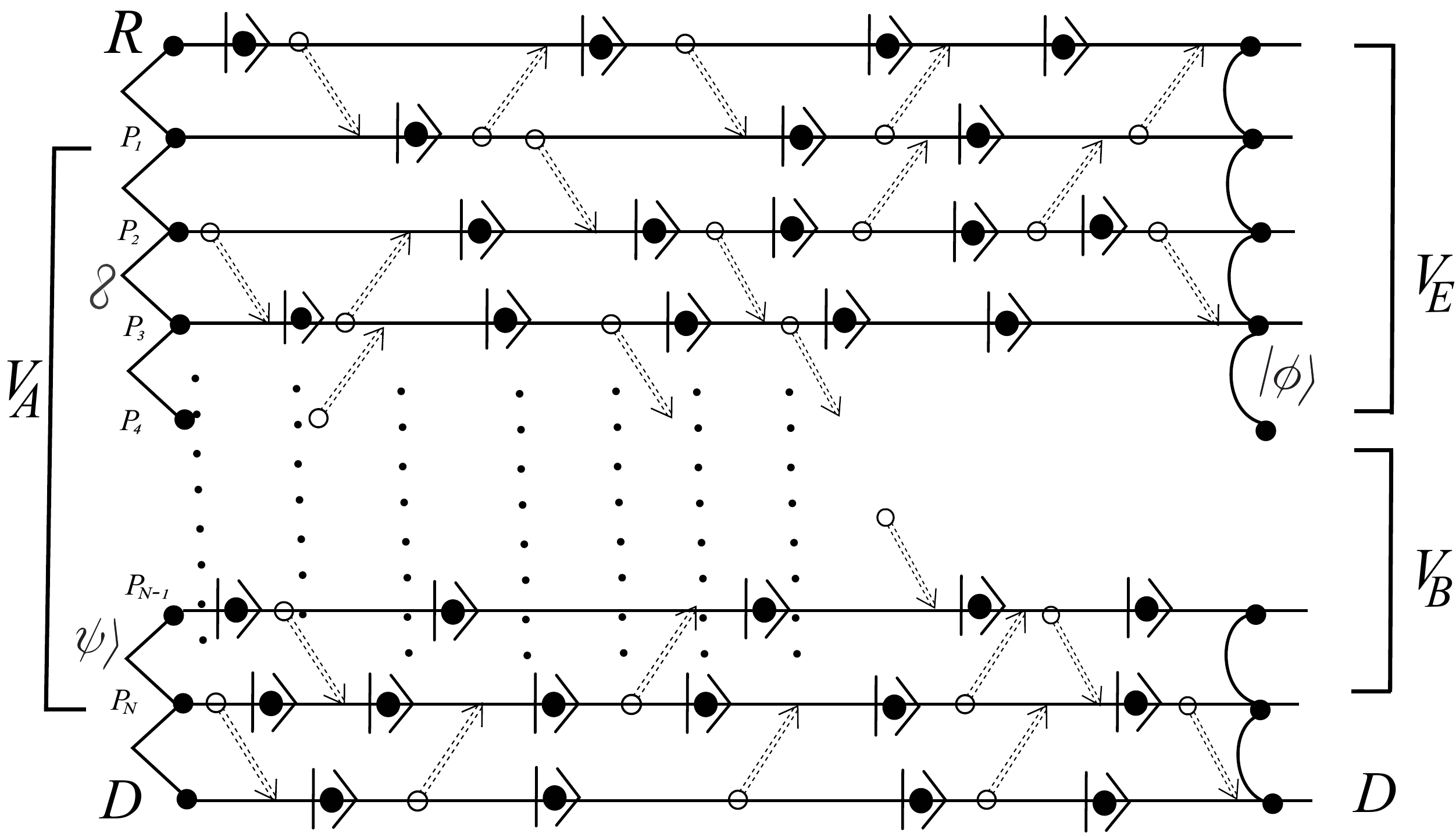}
         \caption{Space-time diagram for quantum information scrambling protocol where node $R$ wants to protect secret quantum information by scrambling across the network. Nodes $P_1$ to $P_{N}$ represent other nodes in the network while $D$ represents the quantum data center, which purifies $P_1$ to $P_{N}$ if necessary. All the local operations are of type II nature as they are from random unitary operations, and the infinity label indicates that quantum entanglement among nodes is assumed to be free. The future light cone of node $R$ can be obtained by following casual paths from $R$. }
     \end{subfigure}
     
        \caption{ Quantum information scrambling protocol in different representations. 
          To get information scrambling, the casual future of the node $R$ must be the entire network. $V_E$ indicates the nodes accessed by a malicious party, while $V_B$ indicates the rest of the network. $D$ denotes the quantum data center node.}
        \label{fig:quantumInformationScramblingProtocol}
\end{figure*}

In this section, we propose a novel protocol for distributed quantum networks inspired by the phenomenon of quantum information scrambling. Additionally, this protocol demonstrates the diagrammatic approach introduced in the earlier section to investigate the quantum information spreading in a network setting. The central concept of the protocol revolves around scrambling quantum information of one particular node $R$ across the quantum network to achieve the following two objectives. Firstly, to reconstruct the quantum information of node $R$, any external malicious entity denoted as $E$ would need to gain access to a substantial subset of the network since the information is not within $R$ anymore. In the most adverse scenario, accomplishing this would necessitate $E$ to breach the entire quantum network. Secondly, even if $E$ were to somehow access the requisite network size, the protocol's construction allows for implementing measures that render the decoding process exponentially complex for $E$, thus making it hard to reconstruct $R$'s information. Therefore, this protocol plays with both information-theoretic and complexity-theoretic arguments to improve the security of a quantum network.

To a certain extent, one could perceive quantum information scrambling as the strongest manifestation of information spreading in a network. All nodes at the end of the protocol are causally related to the node $R$. This feature, that the future of the node $ R $ is the entire network, is a necessary condition for scrambling quantum information. We will now formulate the protocol mathematically. At the start of the protocol, we divide the network $V$ into two subsets, $V_R = \{ R \}$ and $V_A = V \setminus V_R$ described by the quantum state $\rho_{V_A}$. In the next subsection, we will go into the protocol details. For now, suppose the scrambling protocol is implemented successfully. After some time following the protocol, suppose the malicious party $E$ got access to some subset of nodes, which we denote by $V_E$ while the rest of the network is then given by $V_B = V \setminus V_E$. Thus, we have a Hilbert space of $n = |R| + |V_A| = |V_E| + |V_B|$ qubits partitioned as:
\begin{equation}
\label{eq:partition}
    \mathcal{H} = R \otimes V_A = V_E \otimes V_B
\end{equation}
and a unitary map transforming them as:
\begin{equation}
    U_{RV_A}: R \otimes V_A \xrightarrow{} V_E \otimes V_B
\end{equation}
If the initial state is $\rho_R \otimes \rho_{V_A} $, then after the information scrambling protocol $U$, the final quantum state is $\rho '  = U \rho U^\dagger$. The quantum state of a malicious party, $\rho_E$ would be described by tracing out the rest of the network one doesn't have access to $\Tr_{V_{B}} \rho'$. Having done this, we can now precisely formulate the problem as follows: 
\begin{itemize}
    \item What is the minimum size of $|V_E|$ necessary for malicious party E to get access in order to reconstruct the R's quantum information?
    \item Suppose malicious party E has access to this minimum necessary size but doesn't know the unitary $U$, i.e., how the information was scrambled. How many queries does $E$ need to make on different nodes to learn how the information was scrambled?  
\end{itemize}
Interestingly, it turns out that the minimum size $|V_E|$ depends upon the purity, a measure of entanglement entropy, of the system $V_A$. It is possible to change the purity of $V_A$ by entangling it with some other node $D$, which we refer to as a quantum data center. Quantum data centers could serve various purposes in a network. In our context, it primarily functions as a node that, at the protocol's beginning, helps change the purity of $V_A$ by entangling it. The entire setup of the quantum information scrambling protocol, with quantum network graph representation, quantum circuit representation, and space-time diagrammatic representation, is shown in Figure \ref{fig:quantumInformationScramblingProtocol}.

\subsection{Description of the protocol}
\label{sec:DescriptionOfTheProtocol}

In this subsection, we describe how a scrambling unitary $U(t)$ of size $2^{|V|}  \times 2^{|V|}$  can be implemented in a quantum network setting. A detailed analysis will be done in a subsequent section. A trivial way to implement this would be if the node $R$, which wants to scramble the quantum information, has $|V|$ qubits. $R$ performs the $U(t)$ locally and then selects $|P_i|$ qubits randomly and teleport it to each node $P_i$ in a network $V$. However, adopting such a strategy could compromise security since it centralizes all computational activities within a single node. A better method would be implementing the scrambling unitary $U(t)$ in a distributed fashion.  

A scrambling unitary $U(t)$ does not necessarily have any inherent local structure. To account for the locality, for our protocol, we restrict it to be local in neighboring nodes, which are connected by edges. In a quantum network setting, it is reasonable to have all the quantum operations to be local in nodes. A unitary local in neighboring nodes can be made local in nodes by consuming entanglement under LOCC \cite{cacciapuoti2023entanglement}. For example, if a node $A( \in V)$ wants to send an unknown quantum state with dimension $d$ to another node $B( \in V)$, all they need to have is share a bipartite maximally entangled state $\ket{\phi_d}^{AB} := \sum_{i=1}^d \ket{ii}^{AB}/\sqrt{d}$. Thus, we assume that the network is of $G_\phi^\infty$, meaning all the nodes connected by an edge have an infinite supply of maximal EPR pairs at the beginning of the protocol. Let $|P_i|$ be the number of qubits allocated by node $P_i$ for the protocol. It is reasonable to assume that the node $P_i$ can have much more qubits than just $|P_i|$, say at least $|P_i| + \text{max}(|P_{i+1}|, |P_{i-1}|)$. This condition guarantees that node $P_i$ can perform the non-local computation via LOCC, making the entire protocol local in individual nodes.

\begin{figure*}
     \centering
     \begin{subfigure}[b]{0.5\textwidth}
         \centering
         \includegraphics[width=0.4\textwidth]{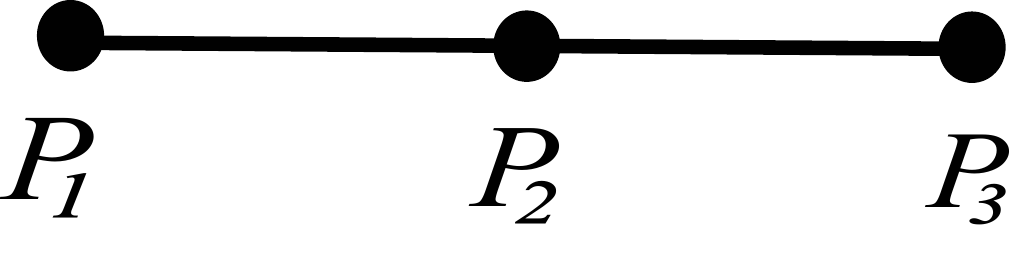}
         \caption{}
     \end{subfigure}
     \hfill
     \begin{subfigure}[b]{0.7\textwidth}
         \centering
         \includegraphics[width=0.6\textwidth]{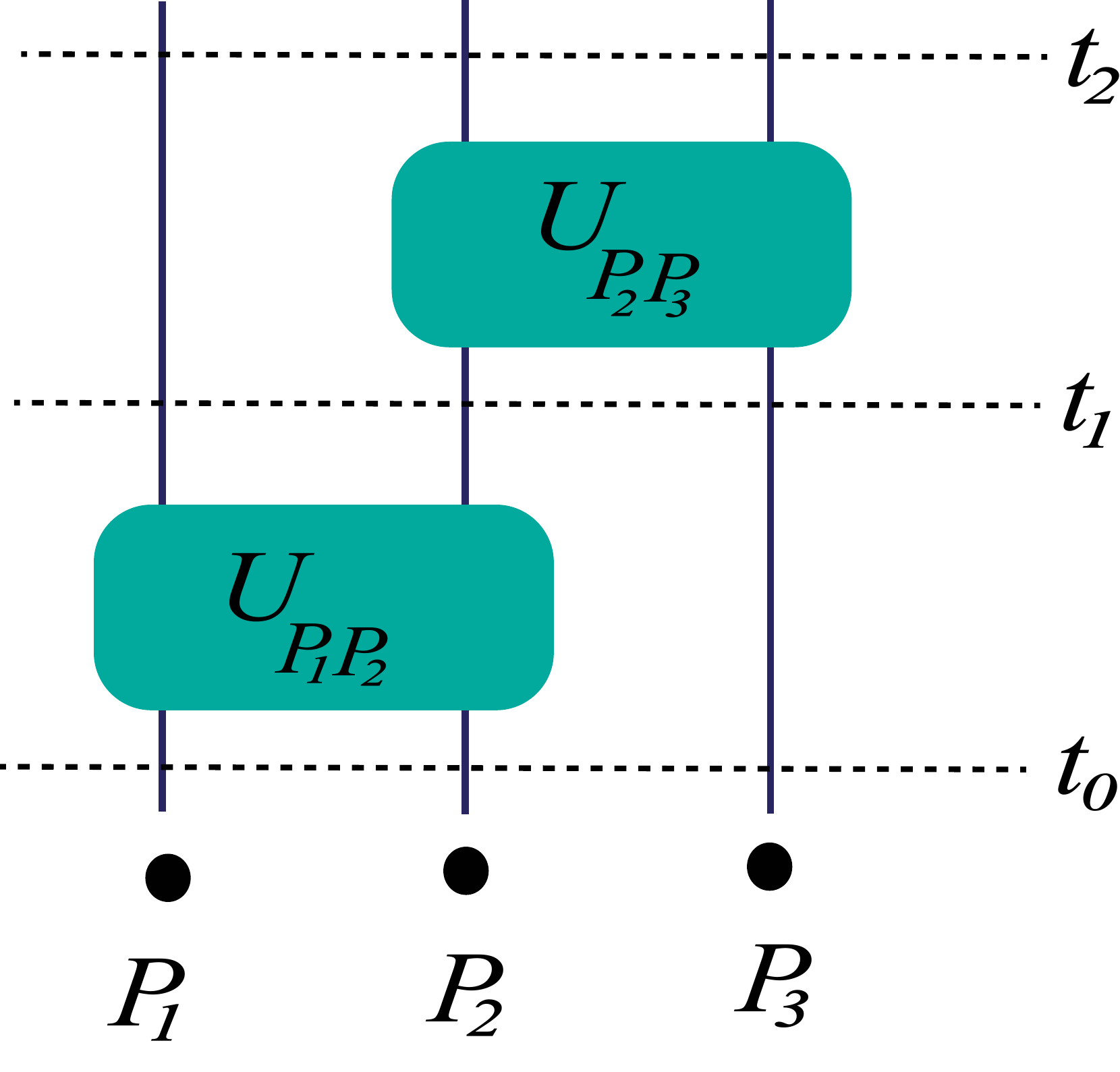}
         \caption{}
     \end{subfigure}
          \hfill
     \begin{subfigure}[b]{0.9\textwidth}
         \centering
         \includegraphics[width=\textwidth]{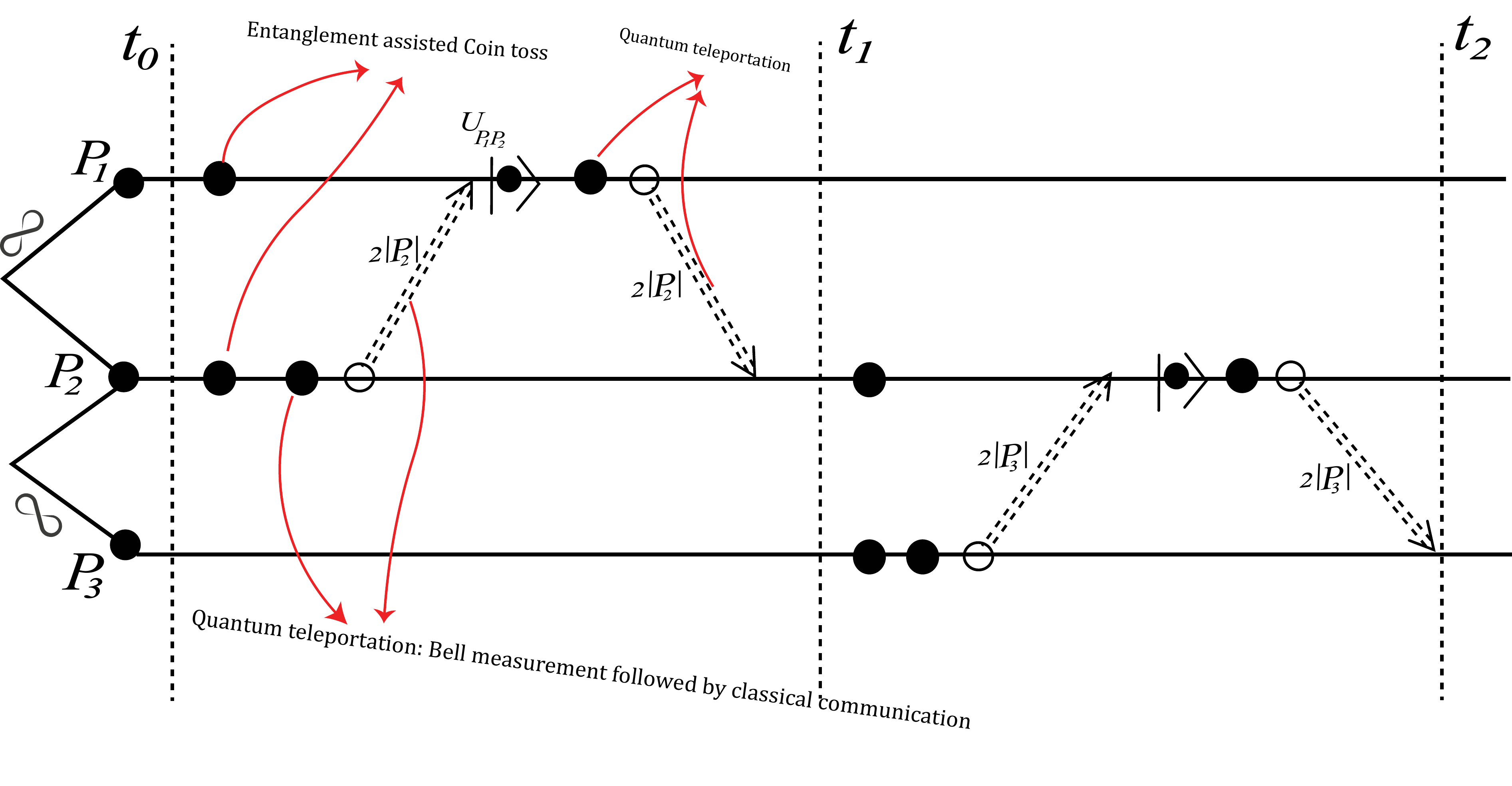}
         \caption{}
     \end{subfigure}

        \caption{(\textbf{a}) A graph $G_\phi^\infty$ with three nodes where $P_1$ $P_2$  and $P_2$ $P_3$ share unlimited EPR pairs. (\textbf{b}) A circuit with three nodes where unitary $U_{P_1P_2}$ is local in nodes $P_1$ and $P_2$ while unitary $U_{P_2P_3}$ is local in nodes $P_2$ and $P_3$. (\textbf{c}) Space-time diagram showing a protocol where the unitary $U_{P_1P_2}$ and $U_{P_2P_3}$ are converted into being local in individual nodes using quantum teleportation.}
        \label{fig:threeNodeProtocol}
\end{figure*}

\begin{figure*}
     \centering
     \begin{subfigure}[b]{0.4\textwidth}
         \centering
         \includegraphics[width = 0.8 \textwidth]{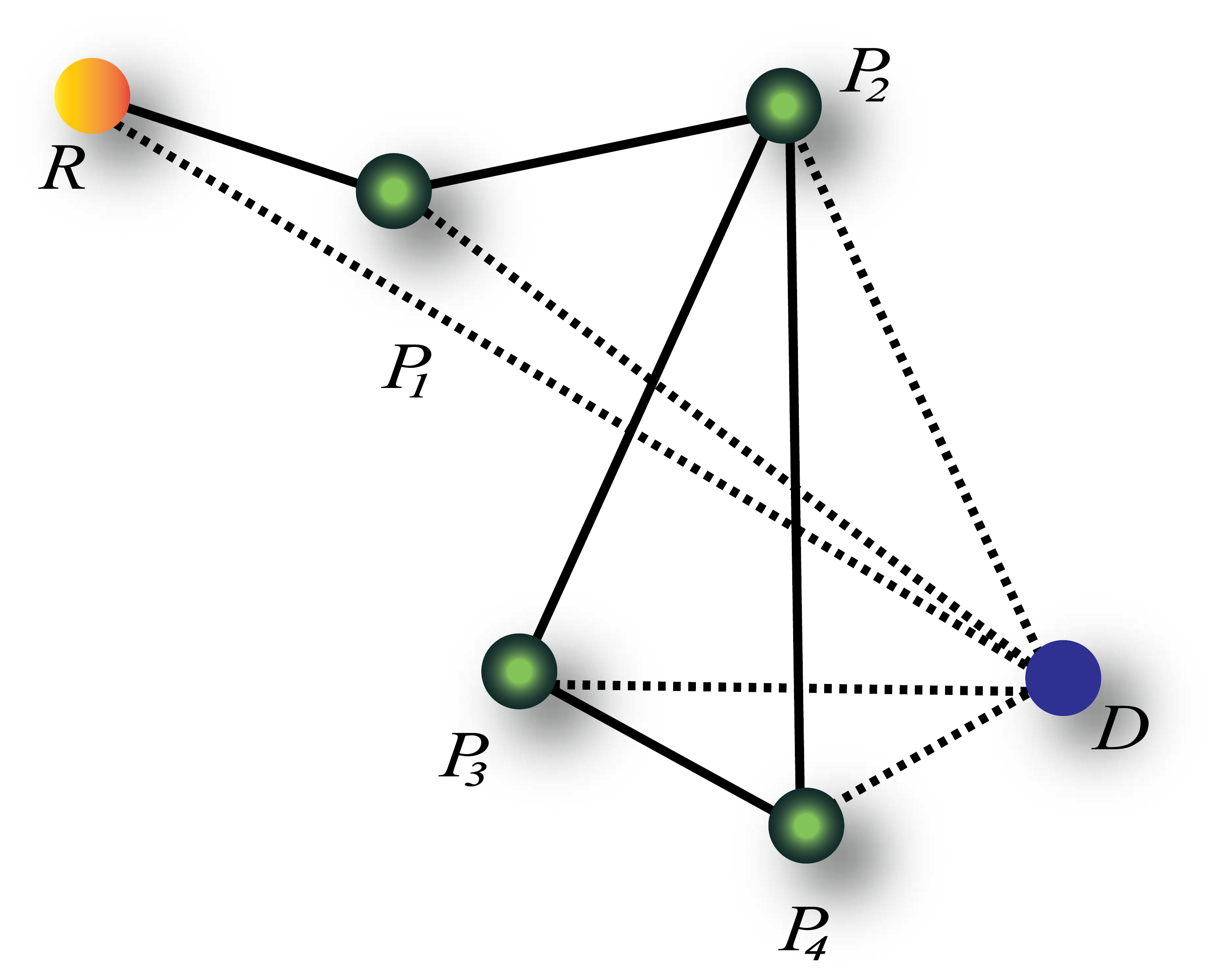}
         \caption{}
     \end{subfigure}
     \hfill
     \begin{subfigure}[b]{0.4\textwidth}
         \centering
         \includegraphics[width =0.8 \textwidth]{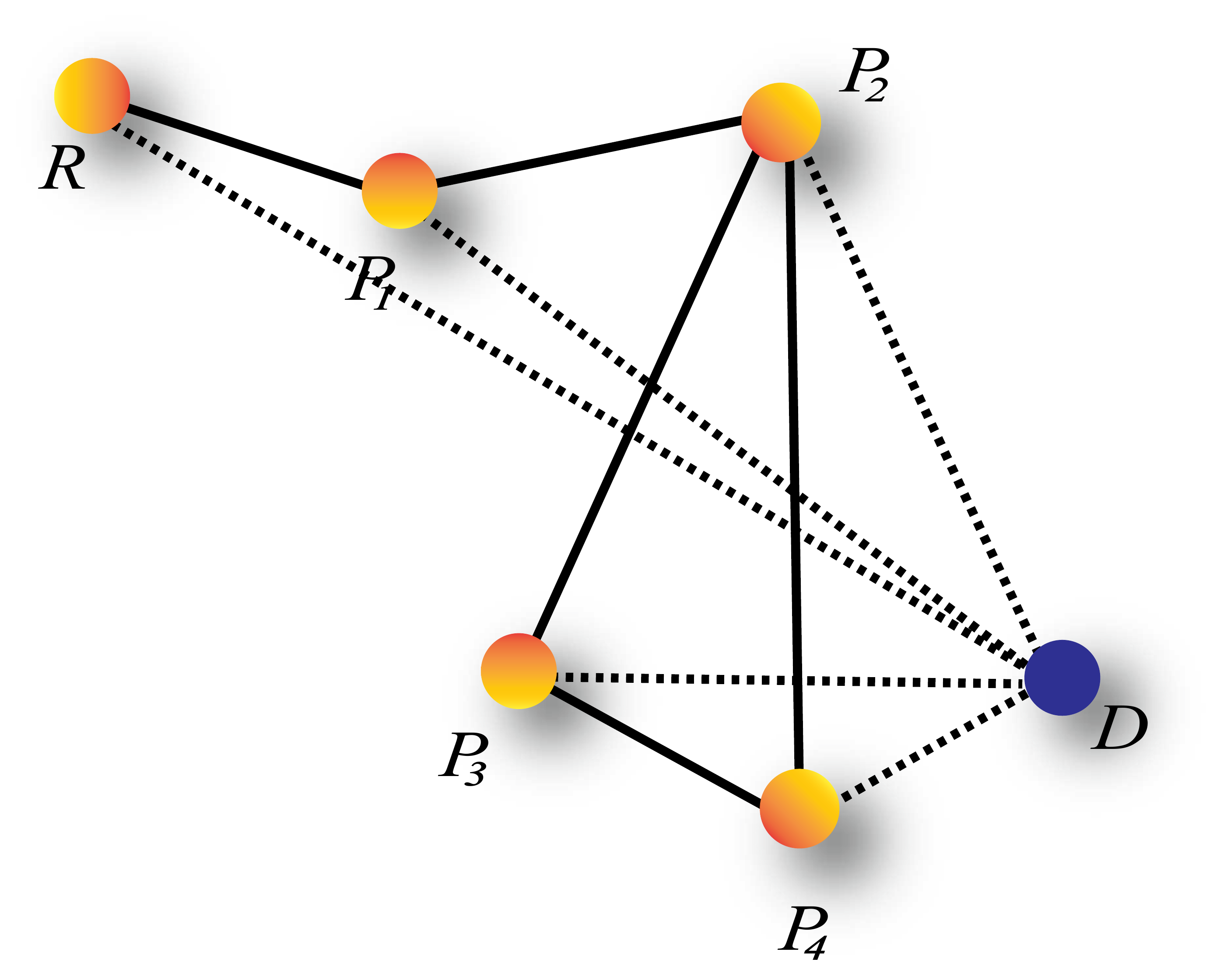}
         \caption{}
     \end{subfigure}
          \hfill
     \begin{subfigure}[b]{0.4\textwidth}
         \centering
         \includegraphics[width =0.8 \textwidth]{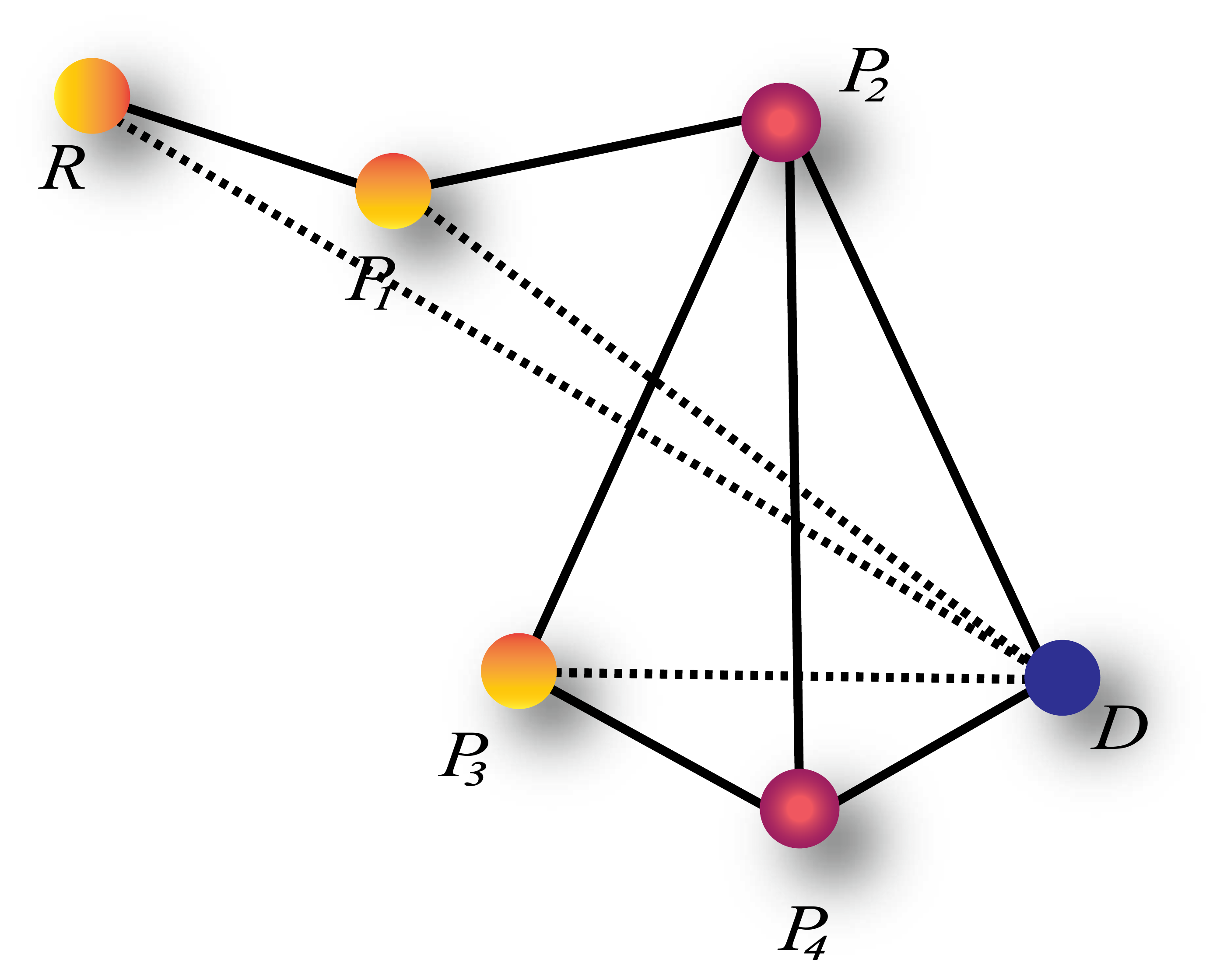}
         \caption{}
     \end{subfigure}
    \hfill
    \begin{subfigure}[b]{0.5\textwidth}
         \centering
         \includegraphics[width =  \textwidth]{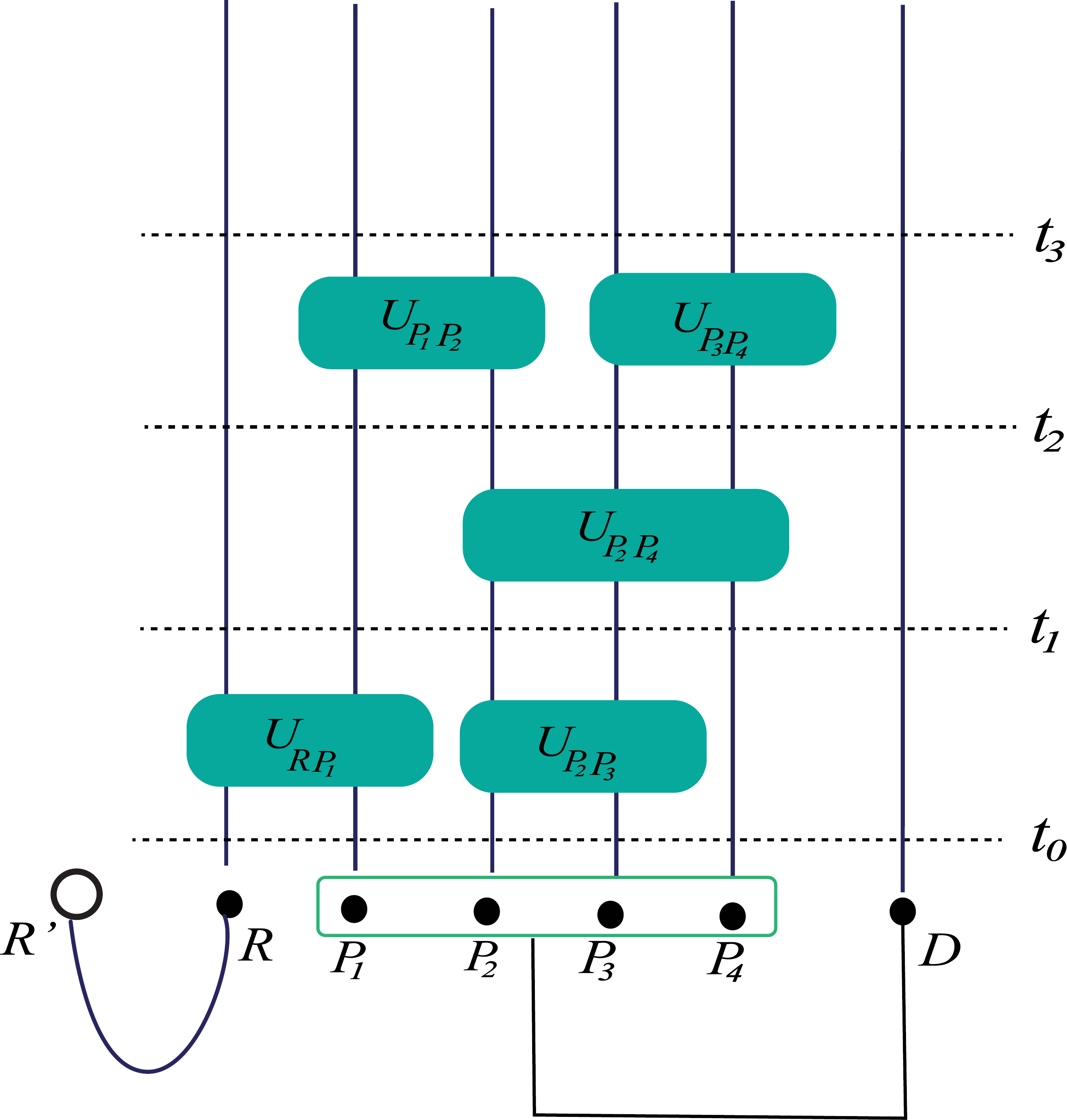}
         \caption{}
     \end{subfigure}
     \hfill
     \begin{subfigure}[b]{0.6\textwidth}
         \centering
         \includegraphics[width = \textwidth]{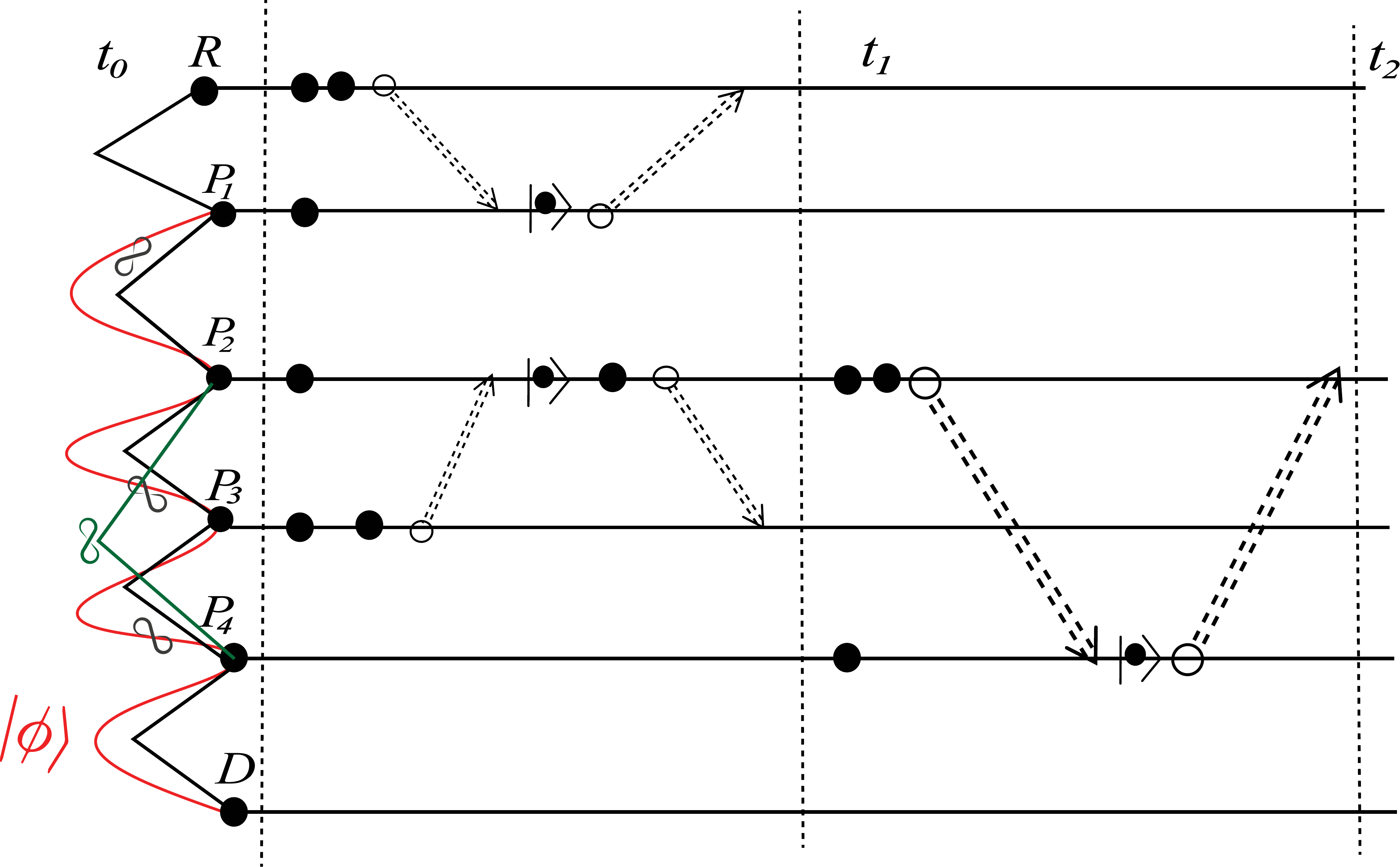}
         \caption{}
     \end{subfigure}
        \caption{(\textbf{a}), (\textbf{b}) and (\textbf{c}) shows the information flow  from a node $R$ in a quantum network $G_\phi^\infty$. $D$ denotes a quantum data center. The dotted line with $D$ indicates the possibility of sharing multiparty quantum resources among the network and data center. (\textbf{d}) shows an example of a possible quantum circuit. $U_{P_2P_4}$ is local in node $P_2$ and $P_4$ as they are connected by an edge as shown in (\textbf{a}).  (\textbf{e}) shows a space-time diagram where all non-local computations of the circuit (\textbf{d}) are made local in nodes using the technique of Figure \ref{fig:threeNodeProtocol}. }
        \label{fig:fiveNodeProtocol}
\end{figure*}

As a simple example, let's consider a network $V \in G_\phi^\infty$ with three nodes $P_1, P_2, \& P_3$ , see Figure \ref{fig:threeNodeProtocol}, with sizes satisfying $|P_1| < |P_2| < |P_3$.
This configuration implies that each node $P_1$, $P_2$ and $P_3$ has at least $|P_1| + |P_2|, | P_2| + |P_3|, |P_2| + |P_3|$ qubits respectively. 
We consider a scrambling circuit with two Haar random unitaries $U_{P_1P_2}$ and $U_{P_2P_3}$, each sized at $2^{|P_1| + |P_2|}$ and $2^{|P_2| + |P_3|}$ respectively, applied sequentially as depicted in Figure \ref{fig:threeNodeProtocol}. Clearly, in the circuit representation, $U_{P_1P_2}$ and $U_{P_2P_3}$ are local in neighboring nodes. We now map this circuit representation into a quantum network using only LOCC protocols. The initial setup involves a $G_\phi^\infty$ network. At time $t_0$, the objective is to implement $U_{P_1P_2}$ at the local node level, either on node $P_1$ or $P_2$. The first step is to flip a coin among $P_1$ and $P_2$, which can be done without any communication cost by consuming entanglement. Furthermore, coin toss introduces randomness in the protocol, making it challenging for potential attackers to breach it. Assuming $P_1$ is chosen, $P_2$ teleports their qubits to $P_1$ through a classical channel by sending $2|P_2|$ classical bits. Since $P_1$ possesses at least $|P_1| + |P_2|$ qubits, $P_1$ can then execute the $U_{P_1P_2}$ unitary of size $2^{|P_1| + |P_2|}$. After the unitary operation, $P_1$ randomly selects $|P_2|$ qubits to teleport them back to $P_2$. This process ensures the local implementation of $U_{P_1P_2}$ at the level of individual nodes. At time $t_1$, the goal is to implement the $U_{P_2P_3}$ unitary. Following a similar approach as with $U_{P_1P_2}$, this can also be achieved locally. The entire protocol is diagrammatically represented in Figure \ref{fig:threeNodeProtocol} where the type II notation is used to represent non-trivial operations such as $U_{P_1P_2}$  and  $U_{P_2P_3}$. Figure \ref{fig:fiveNodeProtocol} shows a more complicated setup.

\subsection{Analysis of the protocol}
In this subsection, we primarily consider the scrambling unitary of Haar scrambling nature, which refers to a unitary $U$ chosen randomly from the Haar measure. We will consider other kinds of scrambling in Section \ref{sec:efficientConstruction}, where we focus on efficiency issues. However, important insights can be drawn from Haar scrambling as we expect the late-time values of entropy and mutual information for any scrambling unitary $U(t)$ to match those of Haar random unitaries.

One can implement Haar scrambling unitary $U(t)$ in the quantum network following the framework outlined in the subsection \ref{sec:DescriptionOfTheProtocol}. Once again, we have a unitary map:
\begin{equation}
    U_{RV_A}: R \otimes V_A \xrightarrow{} V_E \otimes V_B
\end{equation}
where $U_{RV_A}$ is a Haar scrambling unitary local in nodes connected by edges and $R$, $V_A$, $V_B$, and $V_E$ are subnetworks as described in relation \ref{eq:partition}. For the quantum information theoretic analysis, it is often easier to track the quantum information of $R$ in the network by considering an external reference state $R'$ which is perfectly entangled with $R$. Furthermore, we can have a quantum data center $D$ which purifies the subnetwork $V_A$. Thus, the total initial state is $\ket{R'R} \ket{V_A D}$, and the state after the information scrambling protocol is:
\begin{equation}
    \ket{\psi}_{R'DV_EV_B} = U_{RV_A} \otimes I_{R'D}\ket{R'R} \ket{V_AD}
\end{equation}.

\subsubsection{Without quantum data center $D$}

We first consider the case without a quantum data center, i.e., $V_A$ is in a pure state. After applying the  Haar scrambling unitary, it can be shown that (see \cite{Hayden_2007}) the output pure state in $\mathcal{H}_{V_EV_B}$ is likely close to a maximally entangled state if $\frac{|V_E|}{|V_B|} << 1$, more precisely  for the output bipartite Hilbert space $\mathcal{H}_E\otimes \mathcal{H}_{V_B}$, we get
\begin{equation}
\label{eq:bipartiteState}
    \int dU \left|\left| \rho_{V_E}- \frac{Id}{|V_E|} \right|\right|_1 \leq \sqrt{\frac{|V_E|^2 - 1 }{|V_E||V_B| +1 }}
\end{equation}
where $||.||_1$ is the $L_1$ norm and $\rho_{V_E}$ describes the quantum state accessed by malicious party $E$ in the network.
When $|V_B|$ is significantly larger than $|V_E|$, the typical deviation of $\rho_{V_E}$ from maximally mixed state is extremely small. For example, consider the case when the size of $V_B$ exceeds the system in the control of $V_E$ by just ten qubits, then the typical deviation of $V_E$'s quantum state from the maximal mixed state is upper bounded by $2^{-5}$. And, $V_E$ cannot carry any significant information about the scrambled information of $R$. Therefore, as a consequence of equation \ref{eq:bipartiteState}, the malicious party $E$ will need access to at least half the network's size to get any information of $R$.

This result can be intuitively understood as follows. For the pure case, making any arbitrary partition on a network $V$ after information scrambling, if one partition can reconstruct the information of $R$, then the other partition cannot. This follows simply from quantum no-cloning theorem which states that no operation can make copies of an arbitrary unknown quantum state. If both set is able to reconstruct the state $\rho_R$, then no-cloning theorem is violated. This also immediately gives the bound that for $E$ to reconstruct $\rho_R$, $E$ will need to have access to at least half the size of the entire network, even though the nodes which form the set can be completely arbitrary. More concretely, for $V_E$ to be able have any chance  to hack the network, the lower bound on the size of $V_E$ should be $|V_E| = \frac{|V|+1}{2}$, where $|V|$ is the size of the network. 

It is possible to provide a formal security analysis using the language of quantum secret sharing. This follows from the decoupling theorem \cite{Hayden_2008}, which provides a unifying framework for the theory of quantum error correction and many other significant results of quantum Shannon theory. For the decoupling theorem to work, the Haar random unitary encoding must be scrambling, implying the possibility of quantum error correction \cite{r9, Hayden_2008}. On the other hand, the recoverability requirement of a quantum secret-sharing scheme is a sufficient and necessary condition for quantum error correction. Thus, all quantum secret-sharing schemes are also quantum error-correcting codes. A quantum threshold secret sharing scheme is denoted by $((k,n))$ where an arbitrary quantum secret is divided into $n$ parties such that any $k$ or more parties can perfectly reconstruct the secret.
In comparison, any $k-1$ or fewer parties have no information about the secret. Thus, for this pure case, we get a quantum secret sharing scheme similar to $\left(\left(\frac{|V|+1}{2}, |V|\right)\right)$. We plan to further explore this relationship between Haar scrambling and quantum secret sharing in the future. Because Haar scrambling has inherent randomness, it might not be possible to get the exact threshold scheme. Instead, the scheme might be of ramp secret sharing nature. 

\subsubsection{With quantum data center $D$}


We will now discuss the case with quantum data center $D$, which, by entangling with $V_A$ can purify it. Therefore, $V_A$ is in a mixed quantum state as it is part of a pure state $\ket{V_AD}$. The final quantum state after the scrambling protocol is $\ket{\psi}_{R'DV_EV_B} = U_{RV_A} \otimes I_{R'D}\ket{R'R} \ket{V_AD}$. Then, from the conservation of information under unitary evolution, we get:
\begin{equation}
    I(R':R) = I (R':( V \backslash \{D\} = V_BV_E)) + I(R': D)
\end{equation}
where $I(X:Y)$ is the quantum mutual information between $X$ and $Y$.

For the case, when $V_BV_E$ is maximally mixed, $I (R':V_BV_E) = 0$ and all the information about $R$ should now be in $D$ as $ I(R':R) = I(R': D)$. Therefore, when $V_BV_E$ is maximally mixed, even if the malicious party $E$ has access to all the nodes of the network $V$ except for the quantum data center $D$, $E$ can have no information about $R$. From equation \ref{eq:bipartiteState}, this condition is achieved as soon as $|D| > \frac{|V|}{2}$. If the size of $D$ is equal to $\frac{|V|-1}{2}$, $E$ will need to get access to entire subnetwork $V \backslash \{D\}$ to get the information of $R$. If $ 0 < |D| < \frac{|V|-1}{2} $, then in contrast to the pure case of $V_A$, the necessary size of $V_E$ to hack the information of $R$ is greater than $\frac{|V| +1}{2}$. Therefore, by changing the purity (entanglement entropy) of $V_A$ and the size of quantum data center $D$, it is possible to get other security protocols than the one obtained for the pure case.

Furthermore, it is also possible to construct other interesting protocols using a powerful quantum data center. Suppose $V_A$ is maximally entangled with data center $D$. Then, after the information scrambling protocol, $D$ can reconstruct the information of $R$ by just having access to a subnetwork of size $|R|$ from $V \backslash \{D\}$. One can use these features to error-correct quantum nodes in a network or improve security by quickly hiding information if a malicious party $E$ tries to access it. We plan to explore these features in the future.

\subsubsection{Achieving the two criteria}
Now, we will discuss the two criteria we wanted to achieve with the Haar scrambling protocol. The first criterion was that the minimum size of $|V_E|$, necessary for malicious party E to get access to reconstruct the R's quantum information, should be large. In the previous two subsections, we showed that changing the purity of $V_A$ makes it possible to achieve it. Even when $V_A$ is pure, $E$ will need access to half the network to hack the information of $R$. If $E$ is maximally entangled with quantum data center $D$, then $E$ must access the entire subnetwork $V \backslash \{D\}$ to get any information out. We may also already start with $V_A$ being maximally mixed from the start. In this case, $E$ will need to access the entire network $V$ to get the information of $R$. Therefore, by changing the purity of the initial party $V_A$, the information scrambling protocol achieves the first feature we wanted.

The second criterion is related to decoding the information of $R$ that was scrambled across the network. Suppose malicious party E does have access to this minimum necessary size but doesn't know how the information was scrambled. In this case, $E$ must query quantum nodes to learn the scrambling unitary $U(t)$. If $E$ needs to make many queries, it will provide a security advantage in that $E$ cannot decode information of $R$. \cite{leone2023learning} showed that if $U_t$ is a $t$ doped Clifford circuit, it is possible to learn the circuit using only $\mathcal{O}(\text{poly}(n)\exp (t))$ queries where $n$ is the depth of the circuit. The scrambling circuit is a fully chaotic quantum circuit with all local events requiring $t$ gates. Hence, $E$ must make exponential, $\exp(M)$ queries to the network where $M$ is the total number of local events in the protocol. Therefore, the Haar scrambling protocol achieves the second criterion, offering the advantage that even if the information of $R$ is within the grasp of the malicious party E, they cannot decode it.

\subsection{Efficient construction}
\label{sec:efficientConstruction}
Scrambling unitaries do not necessarily have to be perfect Haar unitaries; even simple operators can exhibit excellent scrambling properties. Haar-random and unitary-design states demonstrate nearly maximal entanglement, specifically for $t \geq 2$. This implies that the states obtained from $t$-design circuits \cite{Gross_2007} are information-theoretically indistinguishable from Haar scrambling circuits up to the $t$-th moment \cite{leone2023learning}. Since we employed an information-theoretic argument to derive ramp secret sharing schemes from Haar scrambling, $t$-design circuits are already sufficient and come with the benefit that the whole protocol can be constructed in polynomial time. 
Nevertheless, it is necessary to take caution; otherwise, we may fail to attain the desired second feature. For example, if the scrambling protocol is composed of only Clifford gates, then $E$ can learn the scrambling unitary $U(t)$ with only $\mathcal{O}(\text{poly}(n))$ queries where $n$ is the depth of the protocol. Therefore, it is necessary to have at least $n$ number of $t$ gates for a $n$ depth protocol, such that it will take an exponential, $\mathcal{O}(\exp(n))$, queries by $E$ to decipher the information of $R$. 

Scrambling in a quantum network also depends on the network topology of the underlying graph $G= (V,E)$. Information scrambling can be considered a quantum counterpart of a broadcast algorithm. The time complexity of the scrambling protocol is then lower bounded by the time complexity of the broadcast algorithm, which is of order $\mathcal{O}(\text{Diam}(G))$ for dirty topology, and of order $ \mathcal{O}(|E|) $ for clean topology, where $\text{Diam}(G)$ is the diameter of the graph. One suitable design for the network design is the hyperbolic network design. In Euclidean networks, the number of nodes increases only polynomially as the distance from the center grows. However, in a hyperbolic network, the number of nodes increases exponentially with distance from the center. This allows the spread of information quickly in the network.

\section{Conclusion}
By presenting the diagrammatic visualization approach, motivated by the Hasse diagram and Lamport diagrams, we contribute to the understanding and practical implementation of large-scale quantum networks. These techniques shed light on information flow dynamics, synchronization, error monitoring, resource allocation, and security considerations within the network. In addition, we have proposed a novel quantum information scrambling protocol where malicious parties would need to access a significant fraction of the network to retrieve the information scrambled by some node.

Several intriguing open questions remain to be addressed in our study. Firstly, a deeper investigation into the mathematical structure of space-time diagrams is necessary, with potential insights to be drawn from quantum tensor networks\cite{biamonte2017tensor}, and classical distributed computing \cite{lamport}. Equally important is exploring the practical utility of space-time diagrams across various domains, including synchronization, the construction of fault-tolerant quantum networks, quantum error propagation and correction, and quantum resource tracking. For example, one can explore what else a quantum data center can achieve in terms of constructing fault-tolerant quantum networks and enhancing security in the network. A quantum network's topology shapes a network's information dynamics,  necessitating an in-depth analysis of its relationship with quantum information scrambling protocols. 
Lastly, the ongoing achievements of quantum information theory in many-body systems \cite{Fisher_2023, 1972CMaPh..28..251L}, presents a promising avenue for the development of novel protocols in distributed quantum networks, making it a captivating area for future exploration.

\section*{Acknowledgment}

The research is part of the Munich Quantum Valley, which is supported by the Bavarian state government with funds from the Hightech Agenda Bayern Plus. C.Deppe additionally acknowledge the financial support by the Federal Ministry of Education and Research (BMBF) in the programs with the identification numbers: 16KISK002, 16KISQ028, 16KISQ038, 16K1S1598K, 16KISQ077,  16KISQ093, 16KISR027K.

\bibliographystyle{IEEEtran}
\bibliography{Hasse}

\end{document}